\documentclass[aps,prl,preprint,groupedaddress]{revtex4-1}

\usepackage{graphicx}
\usepackage{dcolumn}
\usepackage{bm}
\usepackage{color}

\usepackage{comment}
\usepackage[geometry]{ifsym}
\begin{document}


\title{Deformation of  Marchenko-Pastur distribution   \\ for the correlated time series
}



\author{Masato Hisakado}
\email{hisakadom@yahoo.co.jp} 
\affiliation{
* Nomura Holdings, Inc., Otemachi 2-2-2, Chiyoda-ku, Tokyo 100-8130, Japan} 
\author{Takuya Kaneko}
\email{tkaneko@icu.ac.jp}
\affiliation{
\dag 
International Christian  University \\
Osawa 3-10-2, Mitaka, Tokyo 181-8585, Japan}


\date{\today}

\begin{abstract}

We study the eigenvalue of the Wishart matrix, which is created from a  time series  with   temporal correlation.
When there is no correlation, the 
eigenvalue  distribution of the  Wishart matrix  is known as the Marchenko-Pastur distribution (MPD) in the double scaling limit.
When there is  temporal correlation, the eigenvalue distribution
converges to the deformed  MPD which has  a longer tail and   higher peak  than the MPD.
Here we discuss the moments of  distribution and  convergence to the deformed MPD for the Gaussian process with a  temporal correlation.
We  show  that the second moment increases as the temporal correlation increases.
When the temporal correlation is the power decay, we  observe  a phenomenon such as  a  phase transition.
When  $\gamma>1/2$ which is the power index of the temporal correlation, the second moment of the distribution is finite and the largest eigenvalue is finite.
On the other hand,   when $\gamma\leq 1/2$, the second moment is infinite and the largest eigenvalue is infinite.
Using  finite scaling analysis, we estimate the critical exponent of the phase transition. 

\hspace{0cm}
\vspace{1cm}
\end{abstract}


\maketitle

\bibliography{basename of .bib file}

\begin{thebibliography}{1}
\bibitem{B} The financial time series data is form Bloomberg.
\end{thebibliography}
\section{I. Introduction}
Random matrix theory (RMT) is a hot topic in   
several fields of physics and mathematics \cite{Meh}. 
It has several applied fields,  such as 
nuclear physics, machine learning, finance, and multiple-input
multiple-output (MIMO) for  wireless communication  \cite{Pot,Pot3,Pot2,Fra,RM,AI}.
  Separating  the noise from the signal  
 is  one of the applications of RTM  in   finance \cite{Pot3,Ikeda1, Ikeda2}.
 For MIMO the  transmission characteristic is ruled by the eigenvalues  of  the  Wishart matrix.
In fact,   the  channel capacity   is calculated by the  distribution of its  eigenvalues.

In this article,  we study  a time series of  financial data.
We created  the Wishart matrix \cite{Wis,Fis}  using these data.
When there is no correlation in the time series  of the data, the distribution of the eigenvalues of the  Wishart matrix  converges to 
the Marchenko-Pastur distribution (MPD) \cite{MP} because  of  Brownian motion.
In the MPD  case   there is an assumption that  the independent  random variables.
We  now discuss the convergence of the  distribution of the eigenvalues of the   Wishart matrix
 when there is  temporal correlation.
Here,  we discuss the exponential and power decay cases.
In MIMO studies, the correlations of the random matrix are important because  we can improve the channel capacity   in correlated fading environments using    large eigenvalues
\cite{cm1,cm2,cm3,cm4}.
The correlation represents the distance between the array  antennas.
In \cite{cm4}  for the exponential decay case, it was conjectured that     the distance between the largest and the smallest eigenvalues of the Wishart matrix increases as the correlation increases  using  numerical simulations.
We show  this conjecture in this article.

In the time series of  financial data,  temporal correlations are   important.
We can observe  temporal correlations in several time series.
The exponential decay corresponds to a short memory,  and the power decay corresponds to
intermediate and long memories  \cite{Long}.
Power decays  are sometimes   observed in  financial time series
as   fractional Brownian motion (fBm)  which includes 
both long and short memories \cite{fbm1, fbm2, fbm3, fbm4}.
In this study,  we use  the  fBm as the power decay case  for the numerical simulations.

In \cite{Ka}, we calculated  the eigenvalue distribution  of the Wishart matrix
and compared the moments  with  the  MPD  
 for 25 financial time series,  which includes
 Crypt currencies, foreign exchange, commodities, government bonds, and stock indexes. 
Note that in the application of the random matrix to  finance,
the time series   of the portfolio is usually used.
In this case,  the Wishart matrix is the correlation matrix of the portfolio.
The  eigenvalue distribution of  the Wishart matrix  does not fit  the MPD well   because of the true correlation in the markets.
In \cite{Pot} and \cite{Pot3}  how to separate  true  correlation and noise using  random matrix theory was introduced.
Here,  we use  the time series of one product.
Therefore, the eigenvalue distribution of  the Wishart matrix fits the  MPD.
In this case,  the Wishart matrix is the correlation matrix for different time series.
We show some examples that are shown in \cite{Ka} in  Table \ref{table2} and 
their  properties   are detailed  in Table \ref{table1}.
We can confirm that  the moments for the   time series,  SOY beans, VIX, and NKY225 fit well with the MPD.
On the other hand,  USD/CAD,  EUR/CHF,  and USD/CNH do  not fit well.
One of the reasons for   fitness for  MPD is the temporal correlations.
In fact, the  times series  of USD/CAD,  EUR/CHF,  and USD/CNH have higher  temporal correlations for the shortest time lag,  and they are 0.2 or more in absolute value as  shown in Table
\ref{Corr}.
We  believe that the  process with  temporal correlations is not adopted by MPD.
What is the distribution instead of  the MPD for these time series with  temporal correlations?
This problem has been  discussed in \cite{Pot2,Sol, Jam}
using  the method of  free probability theory and numerical simulations.
In particular, in \cite{Jam} the case  of fBm was discussed.
In \cite{k1,kw,For,Bai,ben, pca} the perturbation of the  Wishart matrices
was discussed using  correlations.
A  phase transition  is observed in the model.
In the small correlation,
the eigenvalue spectrum is  the same as MPD.
We show the moments of this model in Appendix E.
When the perturbation exceeds the threshold,
we  observe the emergence of a separated  largest eigenvalue.
The   distribution is the same as  MPD.
Hence, the correlation of the perturbation  model  is smaller than our temporal  correlation,   as discussed below.

\begin{table}[h]
\centering
  \caption{Moments for the distribution of  eigenvalues of the financial time series correlation matrix}    and 
 the MPD. $\mu_i$ is the $i$-th moment of the MPD and  the correlation matrix \cite{Ka}.  
 The standard deviations  are shown in parentheses.
  \begin{tabular}{|c|c||c|c|c|c|c|}  \hline
 No. &   Data  & $\mu_2$ & $\mu_3$ & $\mu_4$ & $\mu_5$ &$\mu_6$ \\
 \hline \hline
 & MPD & 1.333  & 2.111 & 3.704 & 6.938 & 13.597 \\
\hline
1&USD/CAD & 1.419 (0.008) & 2.448 (0.028) & 4.716 (0.082) & 9.732 (0.218) & 21.038 (0.568) \\
 \hline 
2&EUR/CHF & 1.430 (0.003) & 2.494 (0.016) & 4.856 (0.055) & 10.113 (0.164) & 22.032 (0.456) \\
 \hline 
3&EUR/GBP & 1.402 (0.003) & 2.385 (0.015) & 4.533 (0.055) & 9.237 (0.180) & 19.738 (0.562)\\
 \hline 
4&SOY & 1.333 (0.003) & 2.111 (0.013) & 3.703 (0.041) & 6.933 (0.122) & 13.569 (0.344)\\
 \hline 
5&VIX & 1.336 (0.007)  & 2.117 (0.027) & 3.713 (0.085) & 6.960 (0.254) & 13.669 (0.736)\\ \hline 
6&NKY 225 & 1.324 (0.049) & 2.081 (0.020) & 3.637 (0.064) & 6.819 (0.186) & 13.433 (0.512)\\
 \hline 
  \end{tabular}
\label{table2}
\end{table}

\begin{table}[tbh]
\caption{Temporal correlation of the shortest time lag in  the financial time series}
\begin{center}
\begin{tabular}{|l|c|c|c|c|c|c|}
\multicolumn{4}{c}{}\\ \hline
No. &1&2&3&4&5&6\\ 
 \hline \hline
Data&
USD/CAD	&	EUR/CHF	&	EUR/GBP	&	\hspace{0.4cm} SOY \hspace{0.4cm}   	&\hspace{0.4cm} 	VIX	\hspace{0.4cm} &	NKY 225\\
\hline 
Corr&
-0.35	&	-0.33	&	-0.32	&	-0.03	&	-0.03	&	-0.08
\\
\hline
\end{tabular}
\label{Corr}
\end{center}
\end{table}

In this study, we study the effects of  temporal correlations of  random variables.
The temporal correlation is the exponential decay and power decay.
When there are temporal correlations,
the eigenvalues  converge to the deformed  MPD.
We  show  that the mean of the deformed MPD does not depend on the correlation  and  that the second moment increases as the temporal correlation increases.
Hence, as the correlation increases, the distribution has a fatter tail and a higher peak.
In  the power decay case,
we    observe a phenomenon such as  a  phase transition.
There are  finite second moment  and  infinite second moment phases.
When  $\gamma>1/2$ which is the power index of the temporal correlation, the second moment of the distribution  and the largest  eigenvalue are  finite.
On the other hand,   when $\gamma\leq 1/2$, the  second moment and the largest  eigenvalue are  infinite.
In fact,     phenomena  such as   phase transition   depend on  temporal correlation \cite{Hisakado6, Hisakado7}. 
 A non-equilibrium phase transition with an order parameter occurs
when the temporal correlation decays according to the  power law \cite{Hisakado6}.
In the case of \cite{Hisakado7} when the power index is less than one,
the  estimator converges slowly. 
They are  in the  non-equilibrium  process and 
 the transition point is $\gamma_c=1$.

The remainder of this paper is organized as follows.
In Section II, we introduce  the time series  and the creation of the Wishart matrix.
In Section III, we 
discuss the distribution of the  deformed 
MPD.
In Section IV, numerical simulations are performed to confirm the deformed MPD.
In section V, we study the phase transition of  the deformed MPD.
Finally, the conclusions are presented in Section VI.

\section{II. Temporal correlation of time series and   Random matrix}

In this section we  introduce  the Wishart  matrix of the time series with  correlation. 
We consider  the time series of a stochastic process,  
$A_t$, to be the   variables at  time $t$.
In the case of  financial data,
we use the historical data of the return $r_t$ as $A_t$.
The return is defined using the market price, $p_t$ as
\[
r_t=\ln p_{t}-\ln p_{t-1}.
\]
Here we  set the normalization,
\[\mbox{E}(A_t)=0,
\]
and 
\[\mbox{V}(A_t)=1.
\]
To introduce the temporal  correlation, let $\{A_t,1\le t \le T\}$ be the time series of the stochastic
  variables of the correlated normal distribution with the following
 $T\times T$ correlation matrix,
\begin{equation}
D_{T-1}=\left(
    \begin{array}{cccc}
     1 &  d_1 &\cdots& d_{T-1} \\
   d_1      &  1 & \ddots&\vdots \\
 \vdots & \ddots   &1&d_1 \\
    d_{T-1}&  \cdots &  d_1  & 1  \\
    \end{array}
  \right),
  \label{matrix}
\end{equation}
which is a parameter for  the time series.
Here the temporal correlation function, $d_t$, is defined as the correlation between $A_s$ and $A_{s+t}$ such that
\begin{equation}
d_t= 
\mbox{Cov}(A_s,A_{t+s}),
\label{Cor}
\end{equation}
in any $s$.
This  shows how  the  previous returns affect this return.
In this article we  consider the exponential decay and power decay cases.
Note that  we use the normalization of $A_t$.
Here  we fold the time series $A_t$, $N$ times,  where $T=N\times L$,
\begin{equation}
A= (\bm{A}_{1}, \bm{A}_{1+L}, \cdots, \bm{A}_{1+(N-1)L}),  
\label{X}
\end{equation}
where $\bm{A}_{\mu}= (A_{\mu},\cdots, A_{\mu+L-1})^T$,
$\mu=1, 1+L, \cdots, 1+(N-1)L$,  is the size $L$ vector  and $ \bm{X}^T$ is the transpose of the vector $\bm{X}$.
$\bm{A}_{\mu}$ is the time series from $\mu$ to $\mu+L-1$.
$A$ is the $L\times N$ matrix.
In the application of the random matrix to the finance,
$\bm{A}_{\mu}$ usually  corresponds to  the different product time series \cite{Pot,Pot3}, but here we use  one product. 

In the matrix form we can rewrite  the $L\times N$ matrix
form
\begin{eqnarray}
A&=&
\left(
    \begin{array}{cccc}
     A_1 &  A_{L+1}  &\cdots& A_{(N-1)L+1} \\
   A_2    &  A_{L+2} & \ddots&\vdots \\
 \vdots & \ddots & \ddots  &A_{NL-1}\\
    A_L&  \cdots &  A_{(N-1)L}  &A_{NL} \\
    \end{array}
  \right).
  \label{EX}
\end{eqnarray}
Note that  here we consider the case $L>>1$.
Hence,
\[
d_i= 
\mbox{Cov}(A_s,A_{i+s})=0,
\]
$i\geq L$.

We consider the relation between the
non correlated random time series 
and  the correlated ones.
Here we consider the $L\times N$ matrix $A_0$ for non correlated time series.
The elements of $A_0$ are i.i.d.
 Here we introduce  the $L\times L$ matrix $\Pi$.
 The relation between the matrix $A$ and $A_0$ is
 \begin{equation}
     A=\sqrt{\Pi}A_0,
     \label{CD}
 \end{equation}
 where
 \begin{equation}
   \Pi=\sqrt{\Pi} \sqrt{\Pi}^T=D_{L-1}.
   \label{pi}
 \end{equation}
 $\sqrt{\Pi}$ is the lower triangle matrix which is created by Cholesky decomposition \cite{k1}.
We show Eq.(\ref{CD}) and Eq.(\ref{pi}) in Appendix F  \ref{AF} in detail.

Next we consider the Wishart matrix of $A$.  
The Wishart matrix is  the correlation matrix $C_{i,j}$ which is a symmetric matrix and the diagonal elements are $1$,
\begin{equation}
C_{i,j}=\frac{1}{L}\sum_{k=1}^{L}A_{(i-1)L+k}A_{(j-1)L+k},
\label{exp}
\end{equation}
where $1\leq i,j\leq N$.
Note that $C_{i,j}=C_{j,i}$, when $i\neq j$ and $C_{i,i}=1$ because of the normalization.
In the matrix form,
\begin{equation}
  C=\frac{1}{L}A^T A=\frac{1}{L}(\sqrt{\Pi}A_0)^T \sqrt{\Pi}A_0, 
\end{equation}
where $A^T$ is the transposed matrix of $A$ and  $C$ is $N \times N$ matrix.

When there is no correlation case, $A=A_0$, the distribution of the eigenvalues of $C$  converges to 
the Marchenko-Pastur distribution (MPS) \cite{MP} in the double scaling limit, $N, L \rightarrow \infty$ with $L/N=Q$.
When $A\neq A_0$,
the distribution of the eigenvalues of $C$  converges  in the double scaling limit, $N, L \rightarrow \infty$ with $L/N=Q$  to
a  distribution, the deformed MPD which is different from MPD.

\section{III.  Convergence to  deformed  Marchenko-Pastur distribution}

In this section we calculate the moments of the 
deformed  MPD.
These  are   new results of this paper and we show   the conjecture of \cite{cm4}. 
Here we consider the case $d_{i}\rightarrow 0$, when $i>>1$.
This  means  that the temporal correlation decays as time goes by.
We  calculate the $k$-th moment of  the eigenvalue distribution of the Wishart matrix, $C$,  
\begin{equation}
\mu_k=\frac{1}{L^{k}N}<\sum_{j=1}^N (x_j)^k>=
\frac{1}{N}<{\rm Tr}(C^k)>,
\end{equation}
where $x_j$ is the eigenvalues of $C$ and
$<>$ means the ensemble average.
The moment of the MPD is in Appendix A.
\subsubsection{i. First moment}
\begin{equation}
    \mu_1=\frac{1}{LN}\sum_{\nu=1}^L\sum_{m=1}^N
    <(A_{\nu m})^2>=1, 
\label{av}
\end{equation}
in the limit of $N, L \rightarrow \infty$ with $L/N=Q$.
Note that it does not depend on the type of  correlation decay.
The mean of the distribution that  does not depend on the temporal correlation is 1.

\subsubsection{ii. Second  moment}
\begin{eqnarray}
    \mu_2&=&\frac{1}{L^2N}\sum_{\nu_1=1}^L\sum_{\nu_2 
 =1}^L\sum_{m_1 =1}^N
    \sum_{m_2 =1}^N< A_{m_1 \nu_1}^T A_{\nu_1 m_2}A_{m_2 \nu_2}^T A_{\nu_2 m_1}> \nonumber \\
    &=&\frac{1}{L^2N}\sum_{\nu_1=1}^L\sum_{\nu_2 =1}^L\sum_{m_1 =1}^N
    \sum_{m_2 =1}^N<A_{\nu_1 m_1} A_{\nu_1 m_2}A_{\nu_2 
 m_2} A_{\nu_2 m_1}>
    =
    1+\frac{1}{Q}+\frac{2}{Q}\sum_{i=1} d_i^2,
\label{mu}
\end{eqnarray}
in the limit of $N, L \rightarrow \infty$ with $L/N=Q$.
Here we assume $d_L\sim0$ in this limit, because 
\[
<A_j,A_{j+L}>=0,
\]
in the limit $L\rightarrow \infty$.
The first and second terms  of Eq.(\ref{mu}) are for   the MPD.
The third term is for the  temporal correlation.
It is the sum of the  cases,  $m_1\neq m_2$ and $\nu_1\neq \nu_2$.
It is 
\begin{eqnarray}
& &\frac{1}{L^2 N}\sum_{\nu_1 \neq\nu_2 }\sum_{m_1\neq m_2 }<A_{\nu_1 m_1} A_{\nu_1 m_2}A_{\nu_2 
 m_2} A_{\nu_2 m_1}>=
 \frac{N(N-1)}{L^2 N}\sum_{|\nu_1-\nu_2|}<A_{\nu_1 m_1}A_{\nu_2 m_1}>^2
 \nonumber \\
 &=&\frac{2N(N-1)L}{L^2 N}\sum_i d_i^2=\frac{2}{Q}\sum_{i=1} d_i^2.
\end{eqnarray}
The second moment of Eq.(\ref{mu}) increases as the temporal correlation increases.
Then  the deformed MPD has a longer tail than  the MPD.
When the second moment is finite, there is the  finite  largest eigenvalue.
On the other hand, the  second moment is infinite, and the  largest eigenvalue is infinite.
We discuss  phenomena  such as  phase transition in  the subsection for the power decay case.

\subsection{A. Exponential Decay case}
We consider the exponential decay case, 
$d_{i}=\mbox{Cov}(A_s,A_{i+s})=r^i,0\le
r\le 1$ and calculate the moments.
$r$ is the effect at $t$ from the past at  $t-1$. 
It is a used temporal correlation for financial time series. 
Here we set 
\begin{equation}
A_{t+1}=rA_t+\sqrt{1-r^2}\xi_t,
\label{A}
\end{equation}
where $\xi_t$ is i.i.d. and we  obtain  the exponential decay  to create a time series with  exponential decay, $<A_{t+1},A_t>=r$.
$\sqrt{\Pi}$ is
\begin{equation}
\sqrt{\Pi}=\left(
    \begin{array}{cccc}
     1 &  0 &\cdots& 0 \\
   r      &  \sqrt{1-r^2} & 0&\vdots \\
 r^2 & r\sqrt{1-r^2} &\sqrt{1-r^2}&0 \\
 \cdots& \cdots & \cdots& \cdots \\
    r^{L-1}&  r^{L-2}\sqrt{1-r^2} &  \cdots  & \sqrt{1-r^2} \\
    \end{array}
  \right),
\end{equation}
 and
\begin{equation}
D_{L-1}=\sqrt{\Pi} \sqrt{\Pi}^T=
\left(
    \begin{array}{cccc}
     1 &  r &\cdots& r^{L-1} \\
   r      &  1 & \ddots&\vdots \\
 \vdots & \ddots   &1&r \\
    r^{L-1}&  \cdots &  r  & 1  \\
    \end{array}
  \right).
\end{equation}

\subsubsection{i. Second  moment}
The second moment is 
\begin{eqnarray}
    \mu_2&=&
    \frac{1}{L^2N}\sum_{\nu_1=1}^L\sum_{\nu_2 
 =1}^L\sum_{m_1 =1}^N
    \sum_{m_2 =1}^N< A_{m_1 \nu_1}^T A_{\nu_1 m_2}A_{m_2 \nu_2}^T A_{\nu_2 m_1})>
    \nonumber \\
    &=&
    1+\frac{1}{Q}+\frac{2}{Q}\frac{r^2}{1-r^2},
\end{eqnarray}
in the limit of $N, L \rightarrow \infty$ with $L/N=Q$.
The first and second terms are for  the  MPD.
The third term is for the deformation for   the correlation.

\subsubsection{ii. Third moment}
The third moment is 
\begin{eqnarray}
    \mu_3&=&\frac{1}{L^3N}\sum_{\nu_1=1}^L\sum_{\nu_2=1}^L\sum_{\nu_3=1}^L\sum_{m_1=1}^N
    \sum_{m_2=1}^N\sum_{m_3=1}^N
    <(A_{m_1 \nu_1}^T A_{\mu_1 n_2}A_{m_2 \nu_2}^T A_{\nu_2 m_3}A_{m_3 \nu_3}^T A_{\nu_3 m_1}) >
    \nonumber \\
    &=&
    1+\frac{3}{Q}+\frac{1}{Q^2}+\frac{6}{Q^2}\frac{r^4}{(1-r^2)^2}+\frac{6}{Q^2}\frac{r^2}{1-r^2}
    +\frac{6}{Q}\frac{r^2}{1-r^2},
\end{eqnarray}
in the limit of $N, L \rightarrow \infty$ with $L/N=Q$.
The first, second and third  terms are of   the  MPD.
The rest  of the  terms are  for the deformation for the correlation.
The fourth term is for the case, $\nu_i\neq\nu_j$ and 
$m_{k}\neq m_{l}$, where $i,j,k,l$ are $1,2, 3$, respectively.
The fifth term is for the case $m_i=m_j\neq m_k$.
It is one  pair of  $m_i$ that  has same values.
The sixth term is for the case $\nu_i=\nu_j \neq \nu_k$.
It is one  pair of  $\nu_i$ which have  the same values.

\subsubsection{iii. Fourth moment}
The fourth moment is
\begin{eqnarray}
    \mu_4
    &=&\frac{1}{L^4N}\sum_{\nu_1,\nu_2,\nu_3,\nu_4=1}^L
    \sum_{m_1,m_2,m_3,m_4=1}^N
    <(A_{m_1 \nu_1}^T A_{\nu_1 m_2}A_{m_2 \nu_2}^T A_{\nu_2 m_3}A_{m_3 \nu_3}^T A_{\nu_3 m_4}
    A_{m_4 \nu_4}^T A_{\nu_4 m_1})>
    \nonumber \\
    &=&
    1+\frac{6}{Q}+\frac{6}{Q^2}+
    \frac{1}{Q^3}
    \nonumber \\
    & &+
    \frac{24}{Q^3}\frac{r^6}{(1-r^2)^3}+\frac{24}{Q^3}\frac{r^4}{(1-r^2)}
    +\frac{24}{Q^3}\frac{r^4}{(1-r^2)^2}
    +\frac{24}{Q^3}\frac{r^2}{(1-r^2)}
    \nonumber \\
    & &+
    \frac{24}{Q^2}\frac{r^4}{(1-r^2)^2}+\frac{12}{Q^2}\frac{r^2}{(1-r^2)}+
    +\frac{24}{Q}\frac{r^2}{(1-r^2)},
\end{eqnarray}
in the limit of $N, L \rightarrow \infty$ with $L/N=Q$.
The first four terms  are for  the MPD.
The rest  of the  terms are  for the deformation for the correlation.
The fifth term is for any $\nu_i$ and $m_j$ which  do not 
 have  the same values.
The sixth and   seventh terms are for one pair which have the same values in $m_i$. 
The eighth term is for two pairs which  have  the same values in $m_i$.
The ninth term is for one pair which has  the same values in $\nu_i$. 
The tenth  term is for one pair which has the e same values in $\nu_i$ and  one pair which has same values in $m_i$.
The eleventh term is for two pairs which have same values in $\nu_i$.

\subsection{B. Power decay case}
In this section we consider the case
 of power decay, $d_i=\mbox{Cov}(A_s,A_{i+s})=1/(i+1)^{\gamma}$ where
$\gamma$ is the power index.
The second moment converges to the finite, when $\gamma>1/2$, 
\begin{eqnarray}
    \mu_2&=&
    \frac{1}{L^2N}\sum_{\mu_1=1}^L\sum_{\mu_2 
 =1}^L\sum_{m_1 =1}^N
    \sum_{m_2 =1}^N<(A_{m_1 \mu_1}^T A_{\mu_1 m_2}A_{m_2 \mu_2}^T A_{\mu_2 m_1})> 
    \nonumber \\
    &=&
    1+\frac{1}{Q}+\frac{2}{Q}\sum_{i=1}^{\infty}\frac{1}{(i+1)^{2\gamma}}
    \nonumber \\
    &<&
 1+\frac{1}{Q}+\frac{2}{Q}\int_{1}^{\infty}\frac{1}{(x+1)^{2\gamma}}dx
    \nonumber \\   
    &=&
    1+\frac{1}{Q}+\frac{2^{2-2\gamma}}{(2\gamma-1)Q}, 
\end{eqnarray}
in the limit of $N, L \rightarrow \infty$ with $L/N=Q$.
On the other hands, $\gamma\leq 1/2$, 
\begin{eqnarray}
    \mu_2&=&
    1+\frac{1}{Q}+\frac{2}{Q}\sum_{i=1}^{\infty}\frac{1}{(i+1)^{2\gamma}}
    \nonumber \\
    &>&
 1+\frac{1}{Q}+\frac{2}{Q}\int_{2}^{\infty}\frac{1}{(x+1)^{2\gamma}}dx
   \nonumber \\   
    &\sim&\lim_{x\rightarrow \infty}
   \frac{2}{(1-2\gamma)Q}x^{-2\gamma+1}. 
\end{eqnarray}
The 
second moment becomes infinite 
and 
the transition point is $\gamma_c=1/2$.

If  $x_1<\infty$  which is the largest eigenvalue and $\sum_i x_i/N\rightarrow 1$ in the limit $N\rightarrow \infty$, then   $\sum_i x_i^2/N<\infty$.
This means finite $\mu_2$.
Conversely if  $x_1\rightarrow \infty$  and $\sum_i x_i/N\rightarrow 1$ in the limit $N\rightarrow \infty$, $\sum_i x_i^2/N \rightarrow \infty$.
This  means infinite $\mu_2$.
Therefore, we can conclude
finite $\mu_2$ corresponds to finite $x_1$ and
infinite $\mu_2$ corresponds to infinite $x_1$.
Therefore,  it is  also a phenomenon like  the phase transition between the finite $x_1$ and  the infinite $x_1$.

\section{IV. Numerical simulations}
In this section  we confirm the conclusions of  section III using  numerical simulations.

\subsection{A. Exponential Decay}

\subsubsection{i. Deformed MPD}
\begin{figure}[htbp]
\begin{tabular}{ccc}
\begin{minipage}{0.33\hsize}
\begin{center}
 \includegraphics[width=5.5cm]{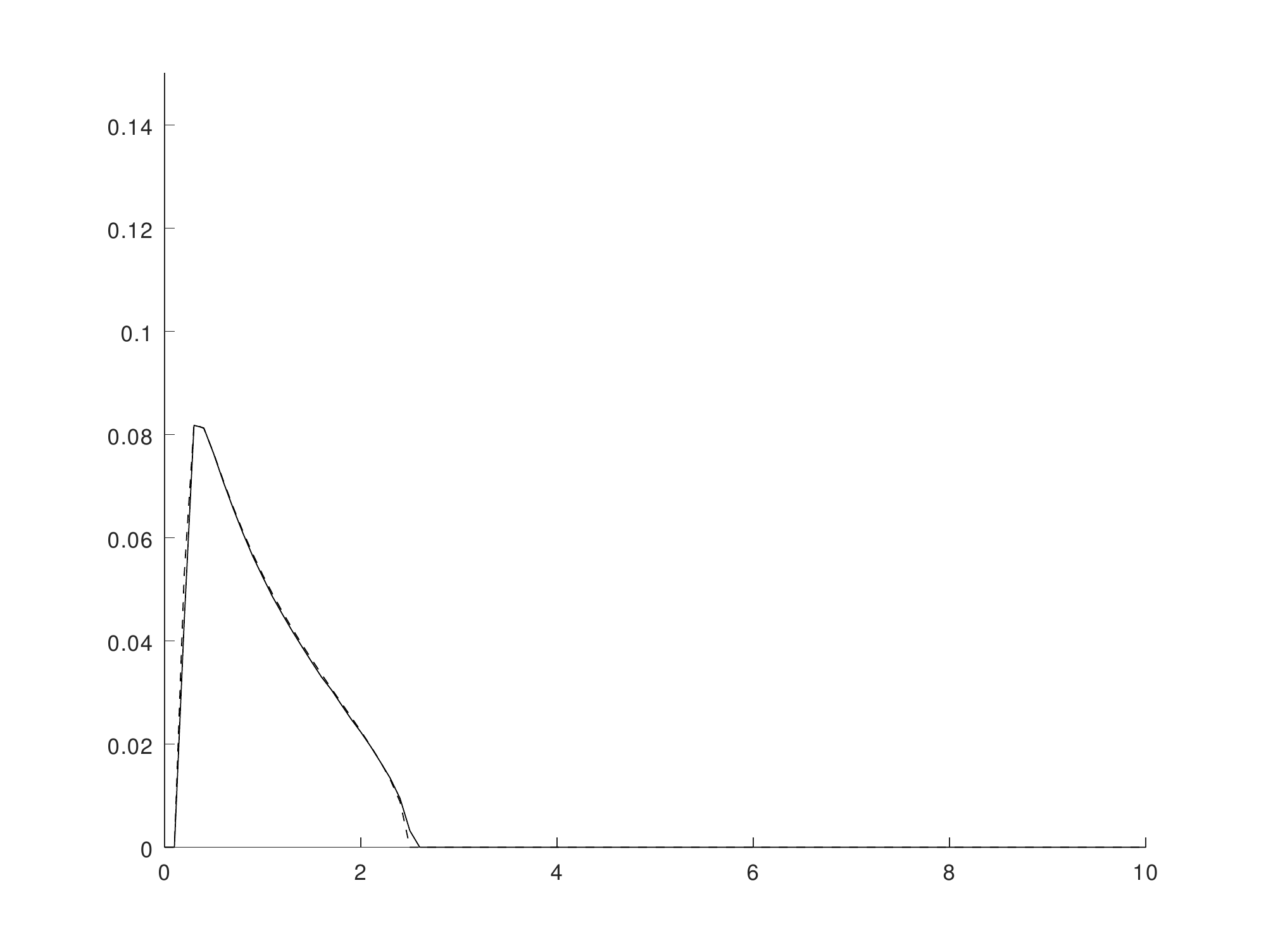} 
\hspace{1.6cm} (a) $r=0.1$
\end{center}
\end{minipage}
 &
\begin{minipage}{0.33\hsize}
\begin{center}
 \includegraphics[width=5.5cm]{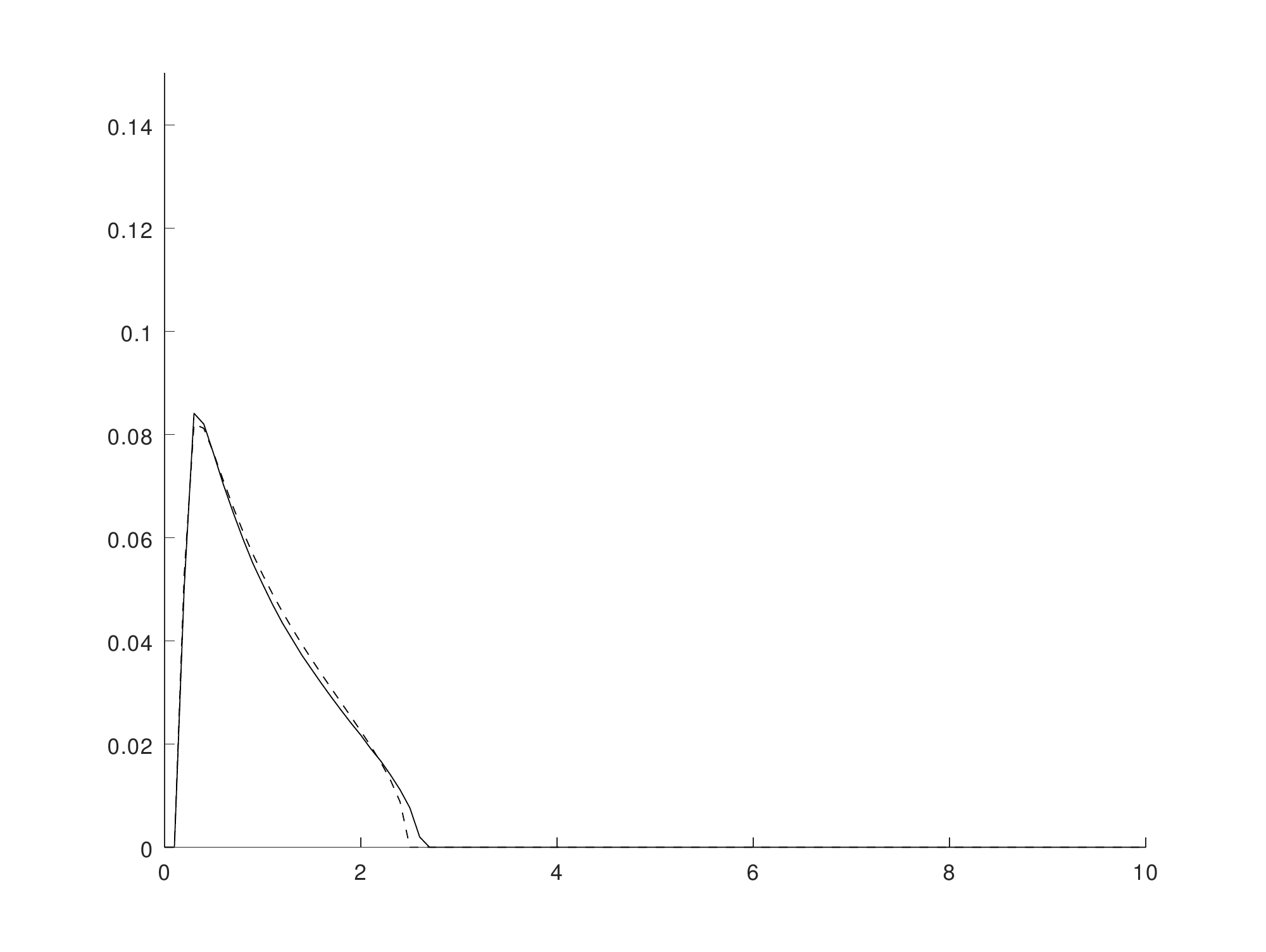} 
\hspace{1.6cm} (b) $r=0.2$
\end{center}
\end{minipage}
 &
 \begin{minipage}{0.33\hsize}
\begin{center}
 \includegraphics[width=5.5cm]{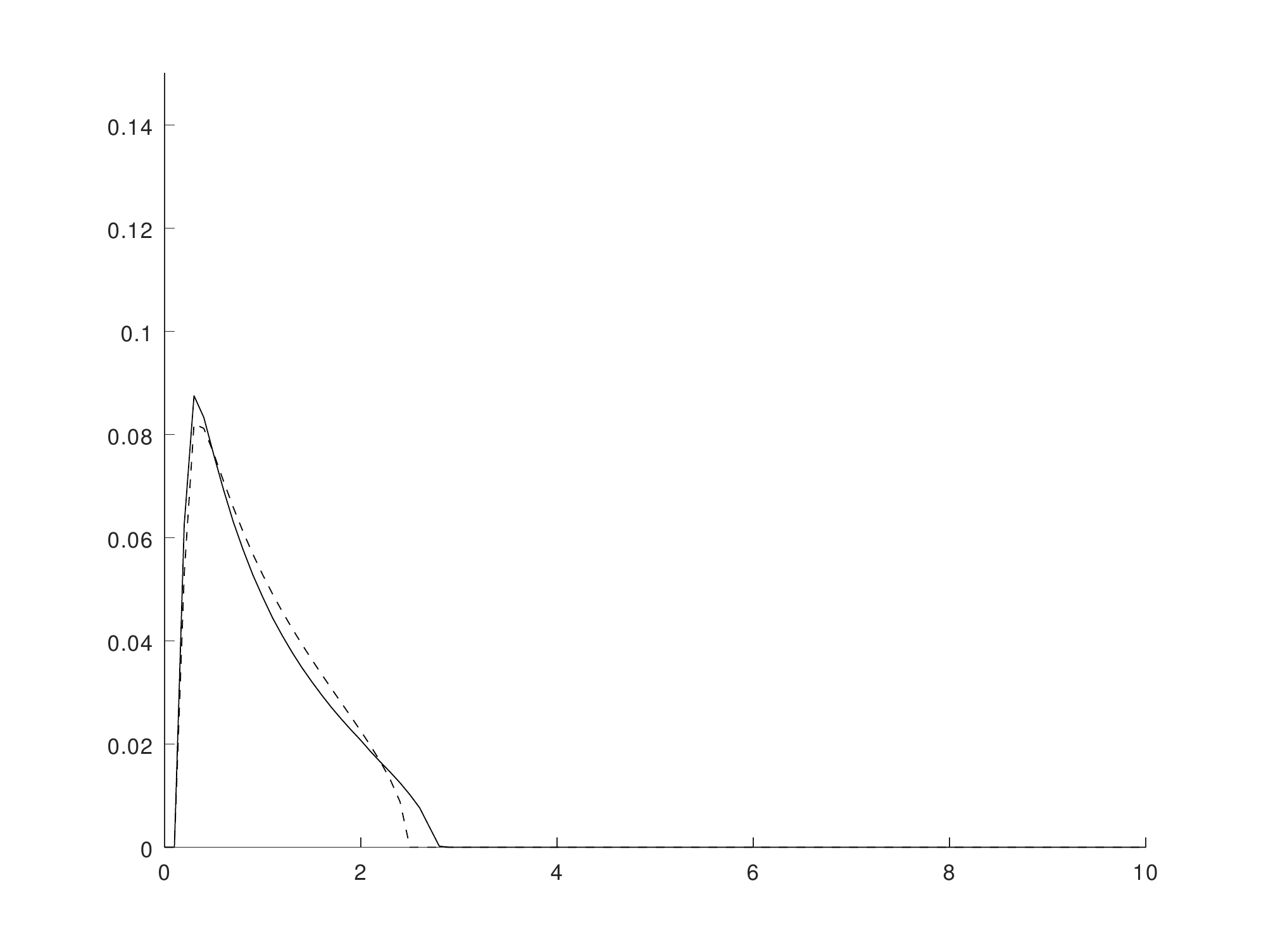} 
\hspace{1.6cm} (c) $r=0.3$
\end{center}
\end{minipage} 
\\
\begin{minipage}{0.33\hsize}
\begin{center}
 \includegraphics[width=5.5cm]{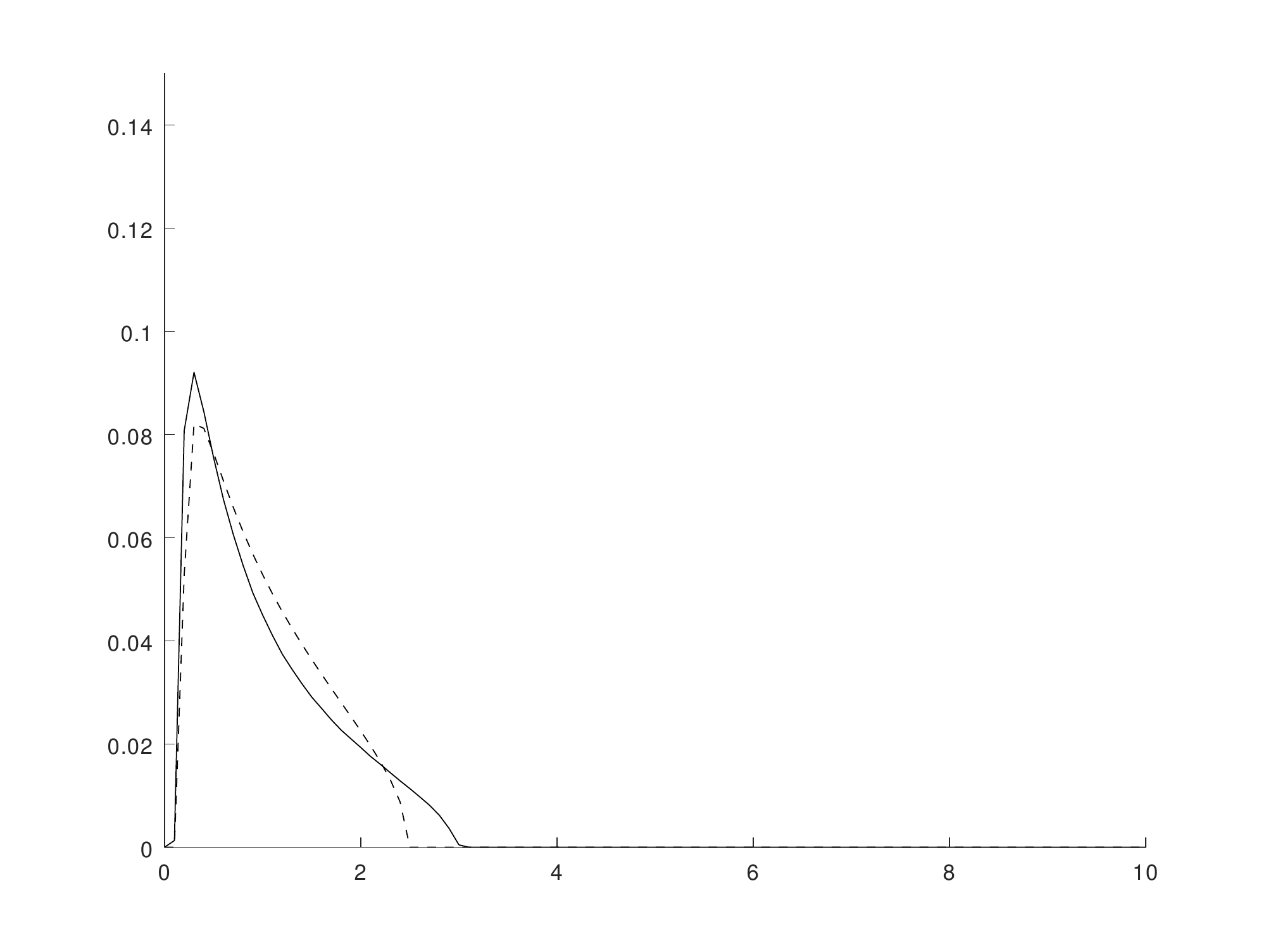} 
\hspace{1.6cm} (d) $r=0.4$
\end{center}
\end{minipage}
 &
\begin{minipage}{0.33\hsize}
\begin{center}
 \includegraphics[width=5.5cm]{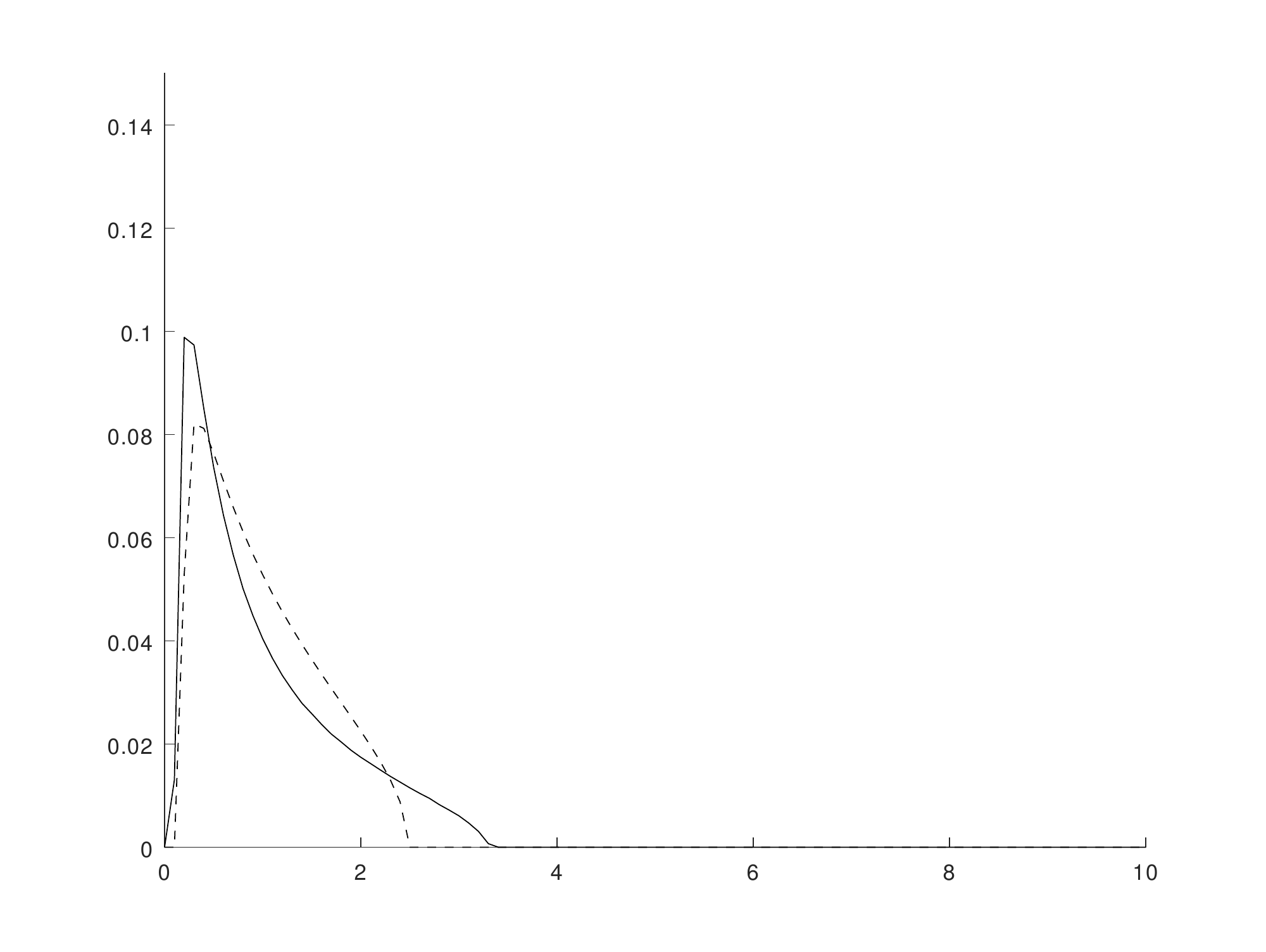} 
\hspace{1.6cm} (e) $r=0.5$
\end{center}
\end{minipage}
 &
 \begin{minipage}{0.33\hsize}
\begin{center}
 \includegraphics[width=5.5cm]{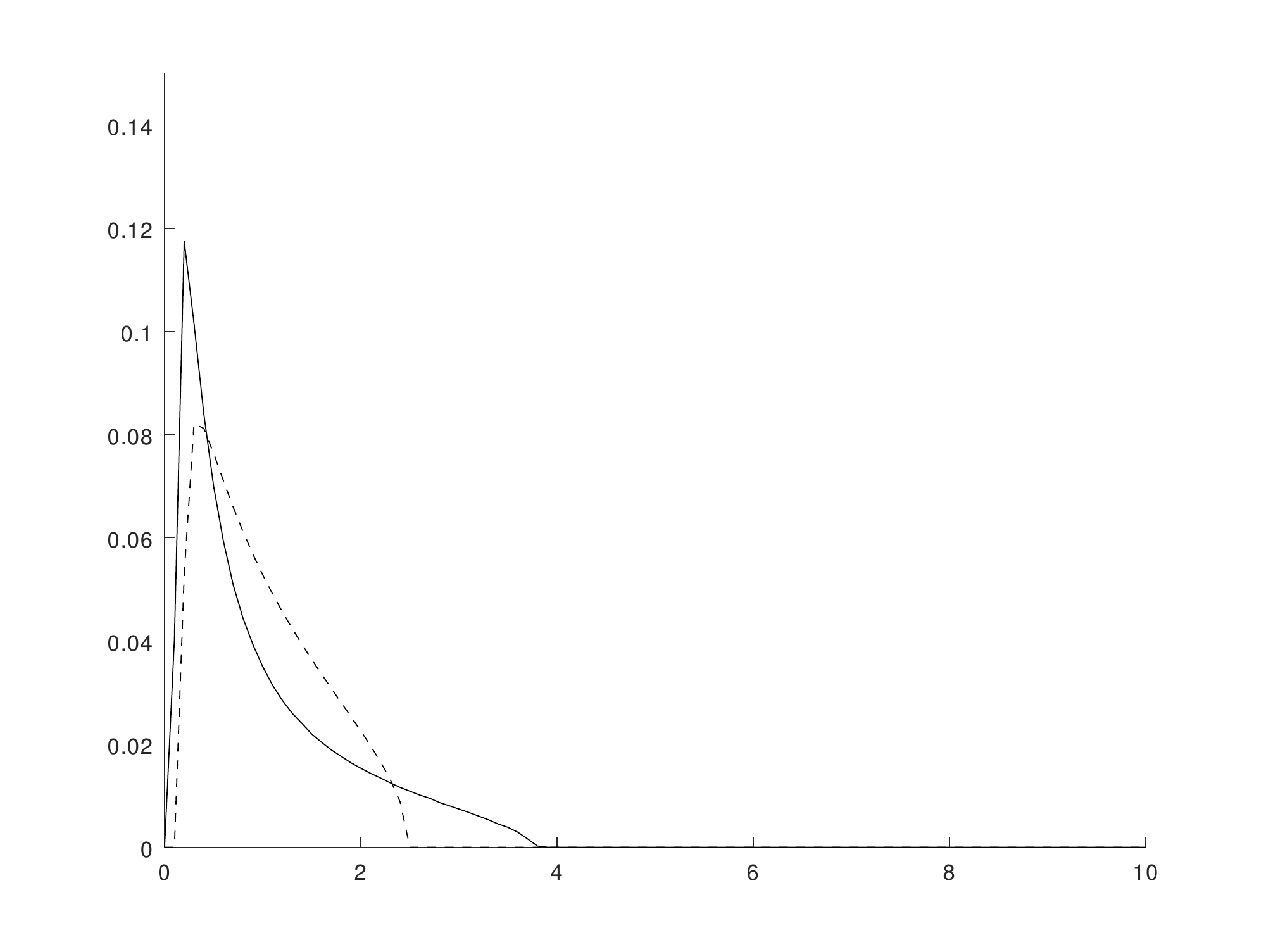} 
\hspace{1.6cm} (f) $r=0.6$
\end{center}
\end{minipage} 
\\
\begin{minipage}{0.33\hsize}
\begin{center}
 \includegraphics[width=5.5cm]{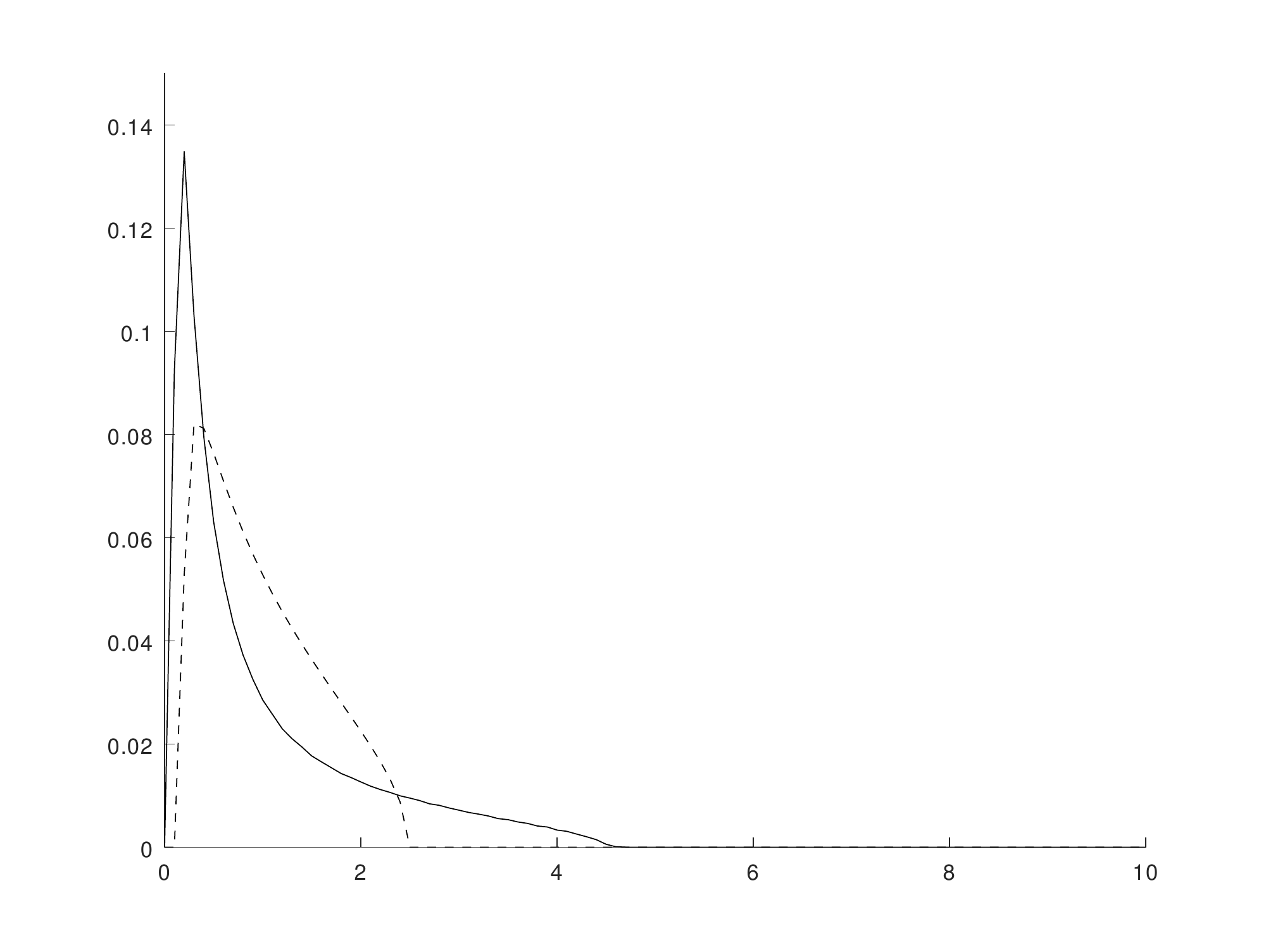} 
\hspace{1.6cm} (g) $r=0.7$
\end{center}
\end{minipage}
 &
\begin{minipage}{0.33\hsize}
\begin{center}
 \includegraphics[width=5.5cm]{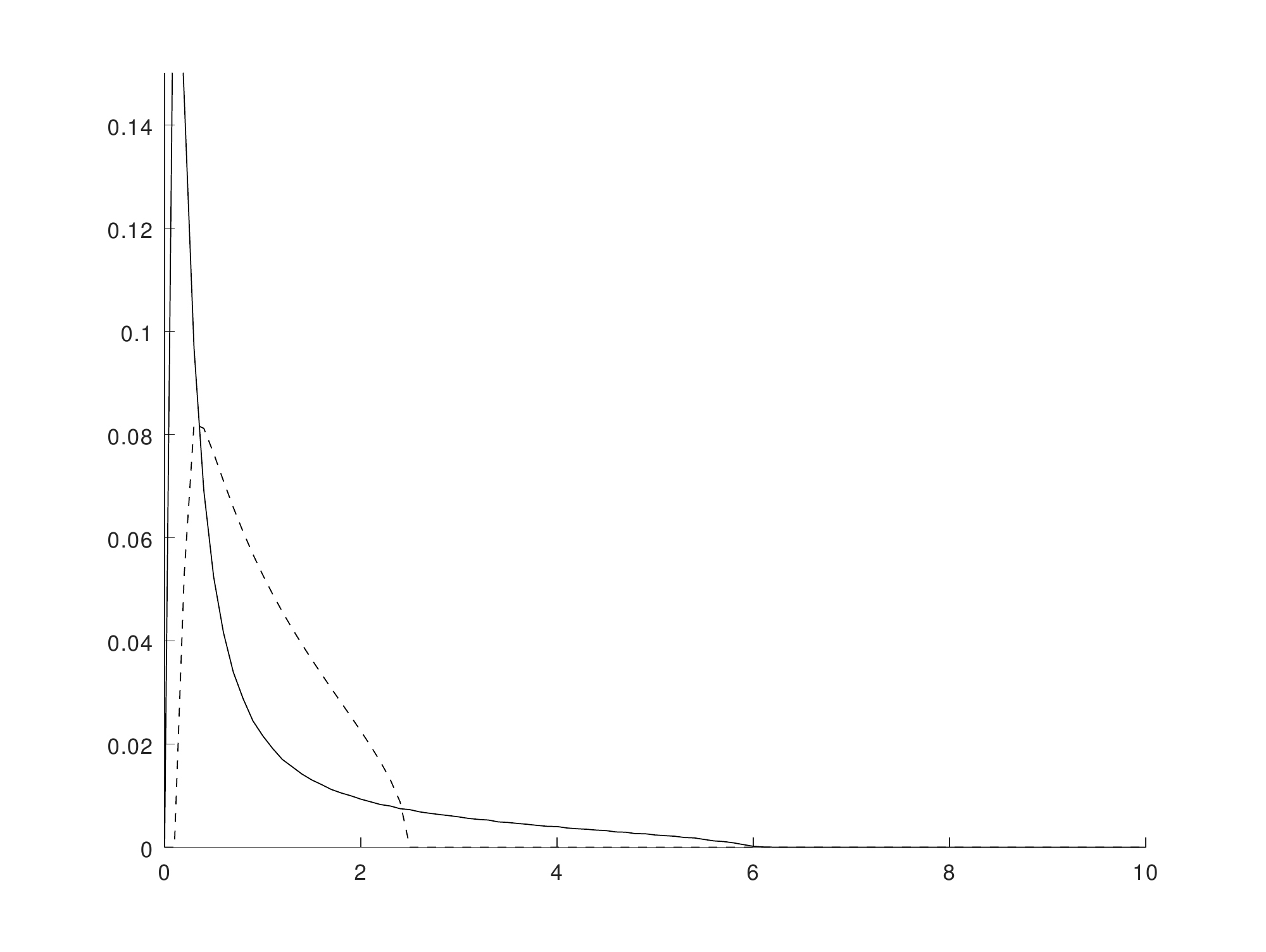} 
\hspace{1.6cm} (h) $r=0.8$
\end{center}
\end{minipage}
 &
 \begin{minipage}{0.33\hsize}
\begin{center}
 \includegraphics[width=5.5cm]{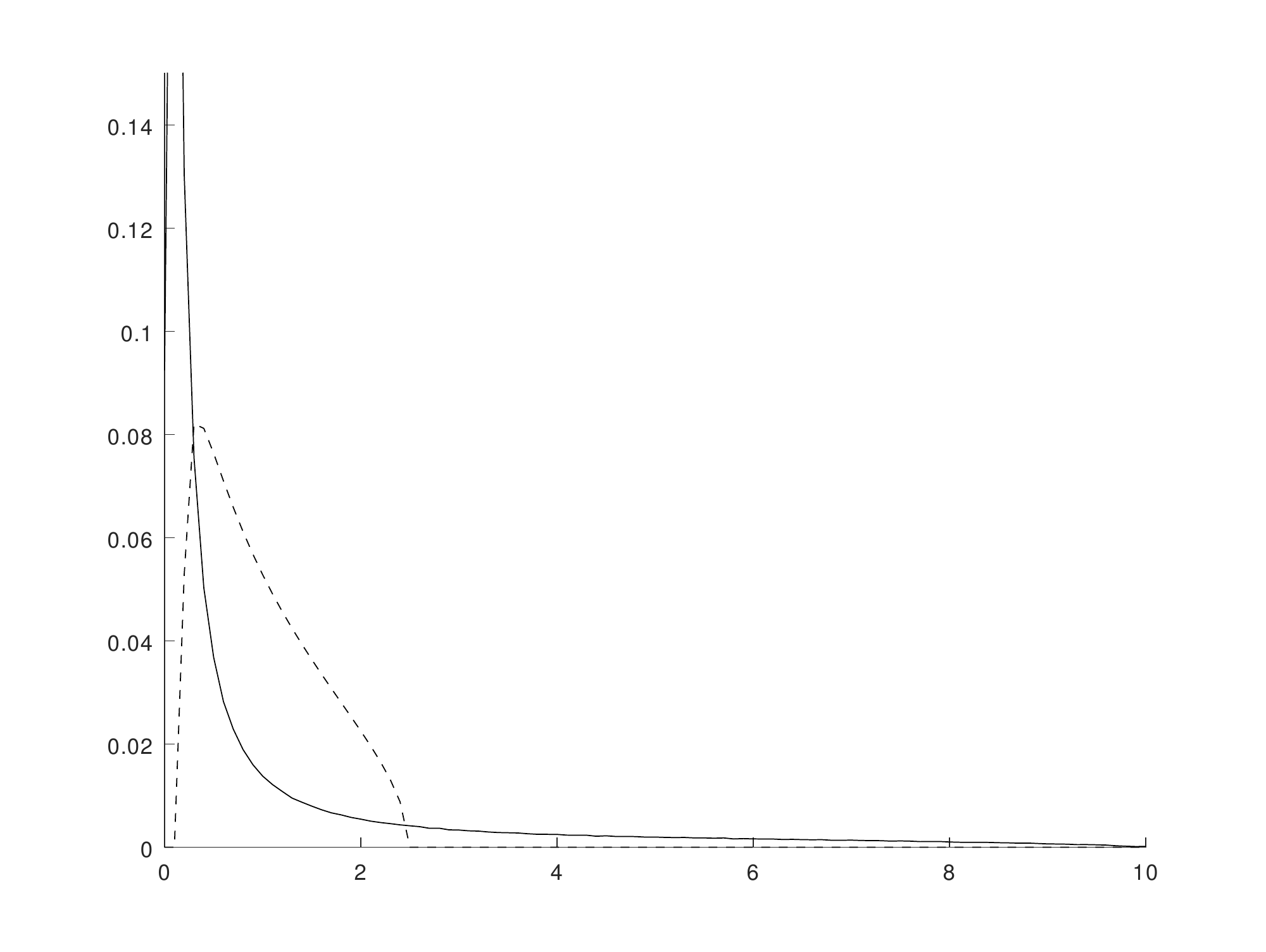} 
\hspace{1.6cm} (i) $r=0.9$
\end{center}
\end{minipage} 
\\
\end{tabular}  
\caption{Plots of  histogram  of the  deformed MPD for $Q=3$ :
(a)$r=0.1$, (b)$r=0.2$, (c)$r=0.3$,
(d)$r=0.4$, (e)$r=0.5$, (f)$r=0.6$,
(g)$r=0.7$, (h)$r=0.8$, (i)$r=0.9$.
The horizontal axis is the eigenvalue and the vertical axis 
is the frequency.
The real line is the distribution with the correlation and the dotted line is the  MPD. We can confirm the fat tail  of the distribution for large $r$ and the mean is constant.
}
\label{MPD}
\end{figure}

First,  we  confirm   the 
deformed MPD.
We  calculated 1000 times each $r$ and 
created a  histogram of the eigenvalues.
The conclusions  are shown in Fig. \ref{MPD}.
We can confirm  that the distributions have  a fat tail and 
 a high peak for large $r$ and the mean of the distribution is  constant.
On the other hand,   for small $r$, the distributions have  almost  the same shape as  the MPD.

\subsubsection{ii. Convergence of the moments}
\begin{figure}[htbp]
\begin{center}
\begin{tabular}{cc}
 \begin{minipage}{0.5\hsize}
\begin{center}
\includegraphics[width=8cm]{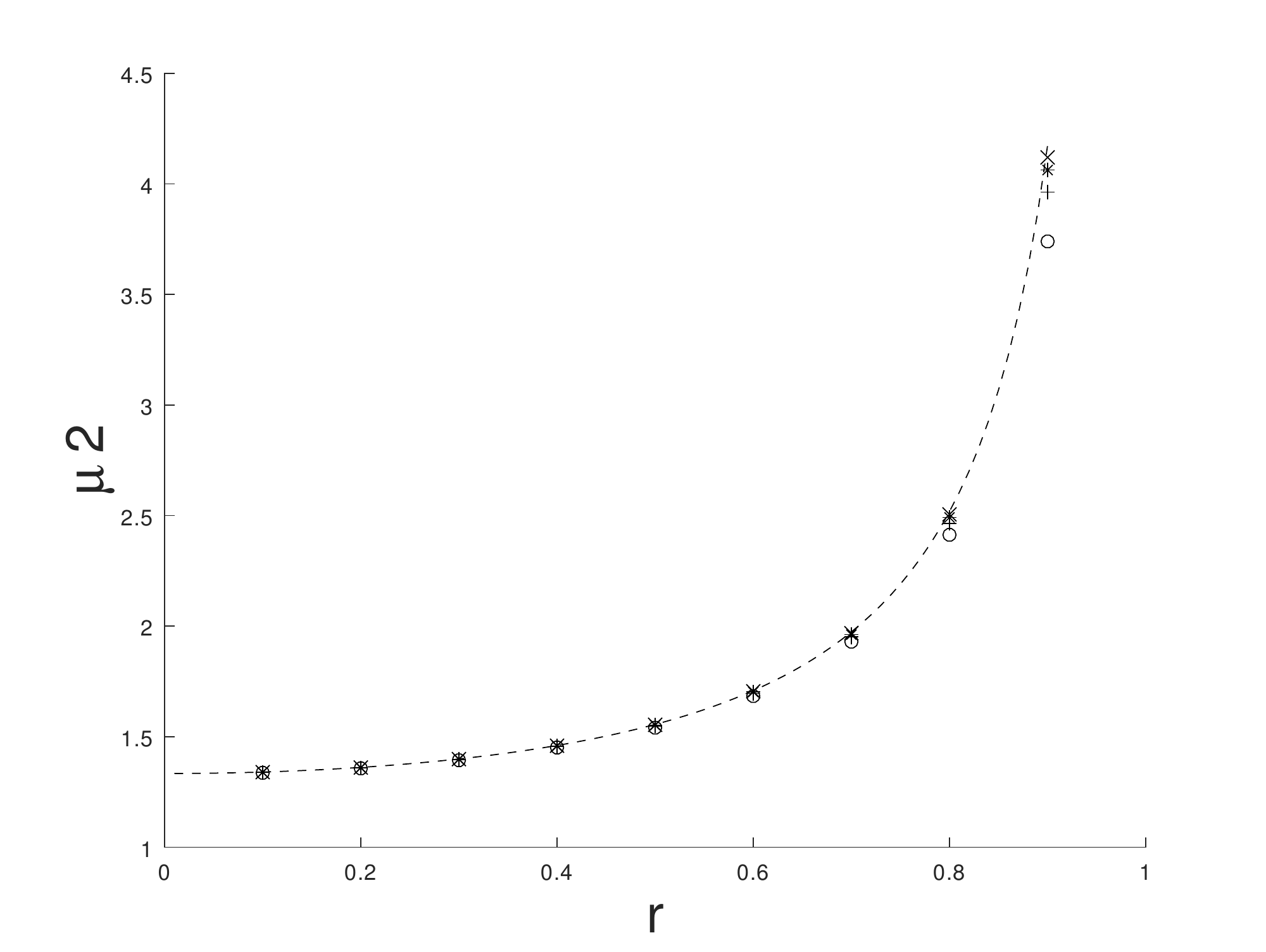}
\hspace{1.6cm} (a)
\end{center}
\end{minipage}
& 
\begin{minipage}{0.5\hsize}
\begin{center}
\includegraphics[width=8cm]{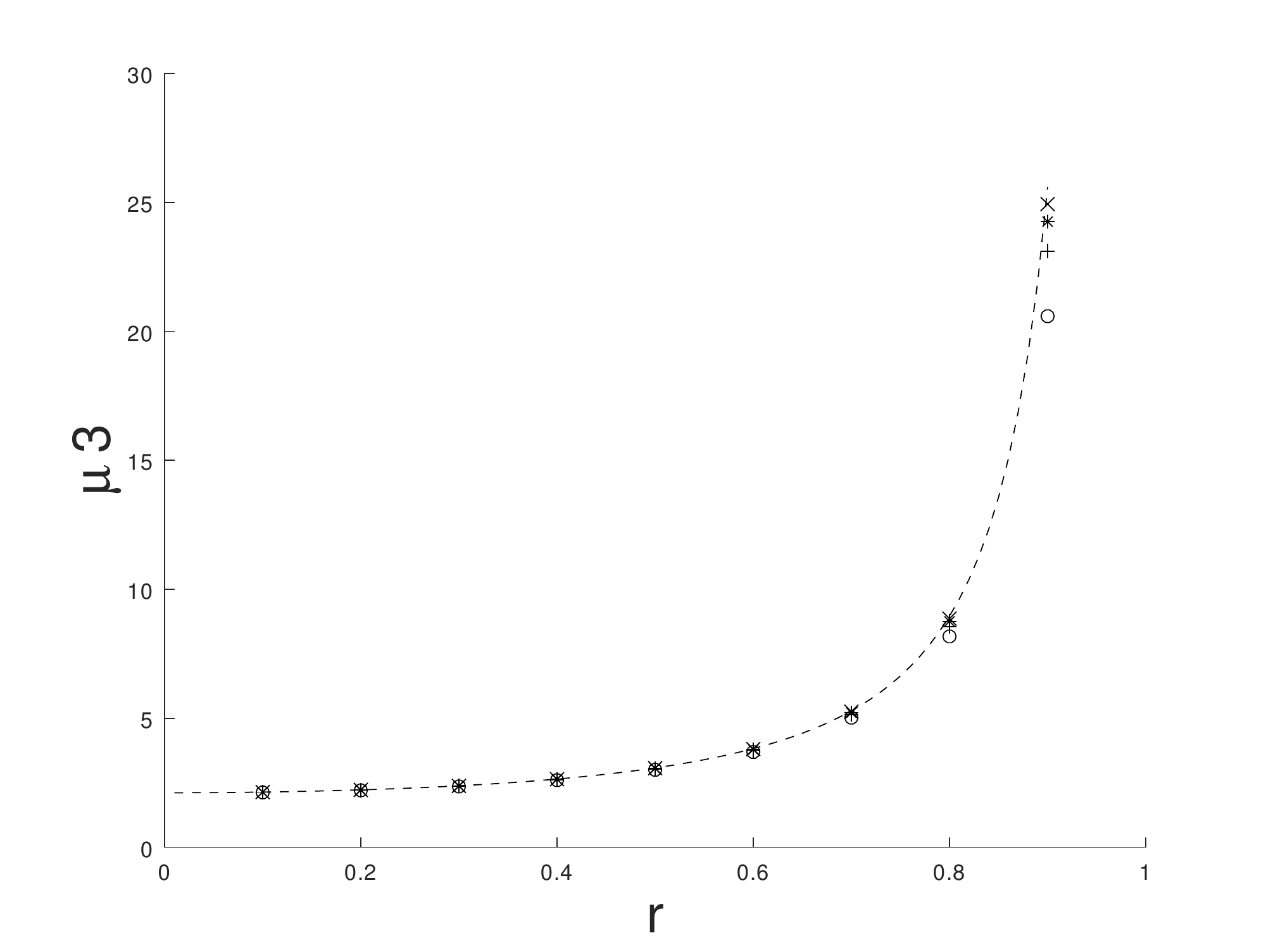}
\hspace{1.6cm} (b) 
\end{center}
\end{minipage}
\end{tabular}
\end{center}
\caption{
The left figure shows the concentration of $\mu_2$ and
the right figure shows the concentration of $\mu_3$.
The horizontal axis is $r$ and the vertical axis is the second and third moments.
Line styles: "o", "+", "*", "x", and the dotted  lines indicate $N=64, 128, 256, 512$ and theoretical moments, respectively. 
As the matrix size increases, the moments converge to
the theoretical moments.
Curves are averages of 1,000  replications
and the moments are the averages.
}
\label{concent}
\end{figure}

Next, we confirm the convergence of the distributions for the exponential decay case.
In Fig.\ref{concent} we show the convergence of the moments, $\mu_2$ and  $\mu_3$.
The horizontal axis is $r$ and the vertical axis represents  the second and third moments.
As the matrix size increases,   the moments converge to
the theoretical ones.
In  Appendix C,   we present the conclusions of  numerical simulations in detail.

\subsection{B. Fractional Brownian motion} 
Next, we confirm  the fractional Brownian motion (fBm)
 case which has   power decay temporal correlations and  fBm is observed in the financial time series.
The relation of the indexes is 
\begin{equation}
    2-2H=\gamma,
\end{equation}
where $H$ is    the Hurst index for  fBm and $\gamma$ is the power index.
Hence, the transition point is,  $H_c=3/4$ which corresponds 
to $\gamma_c=1/2$.
We explain  fBm in Appendix B.

\subsubsection{i. Deformed MPD}
\begin{figure}[htbp]
\begin{tabular}{ccc}
\begin{minipage}{0.33\hsize}
\begin{center}
 \includegraphics[width=5.5cm]{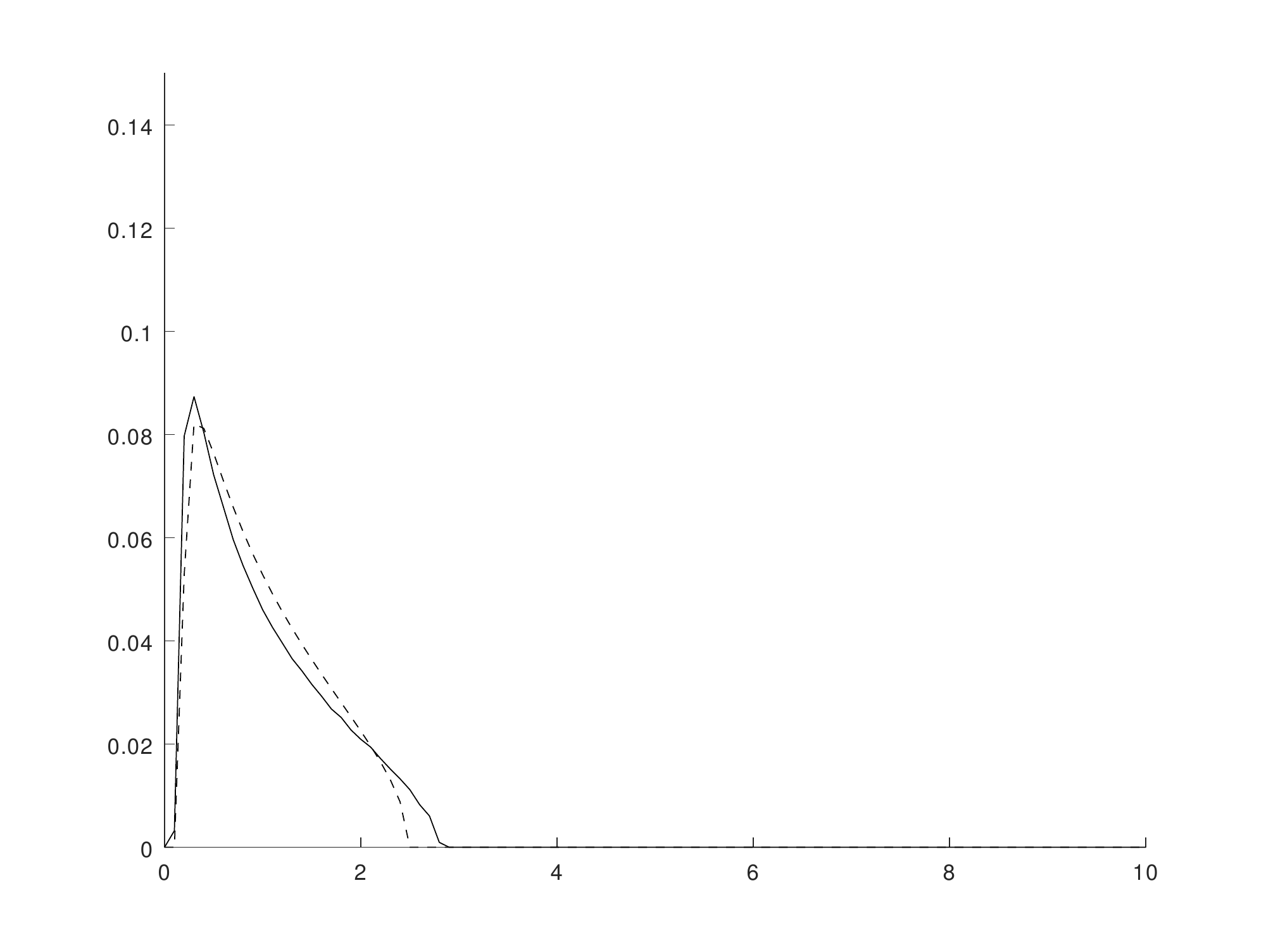} 
\hspace{1.6cm} (a) $H=0.1$
\end{center}
\end{minipage}
 &
\begin{minipage}{0.33\hsize}
\begin{center}
 \includegraphics[width=5.5cm]{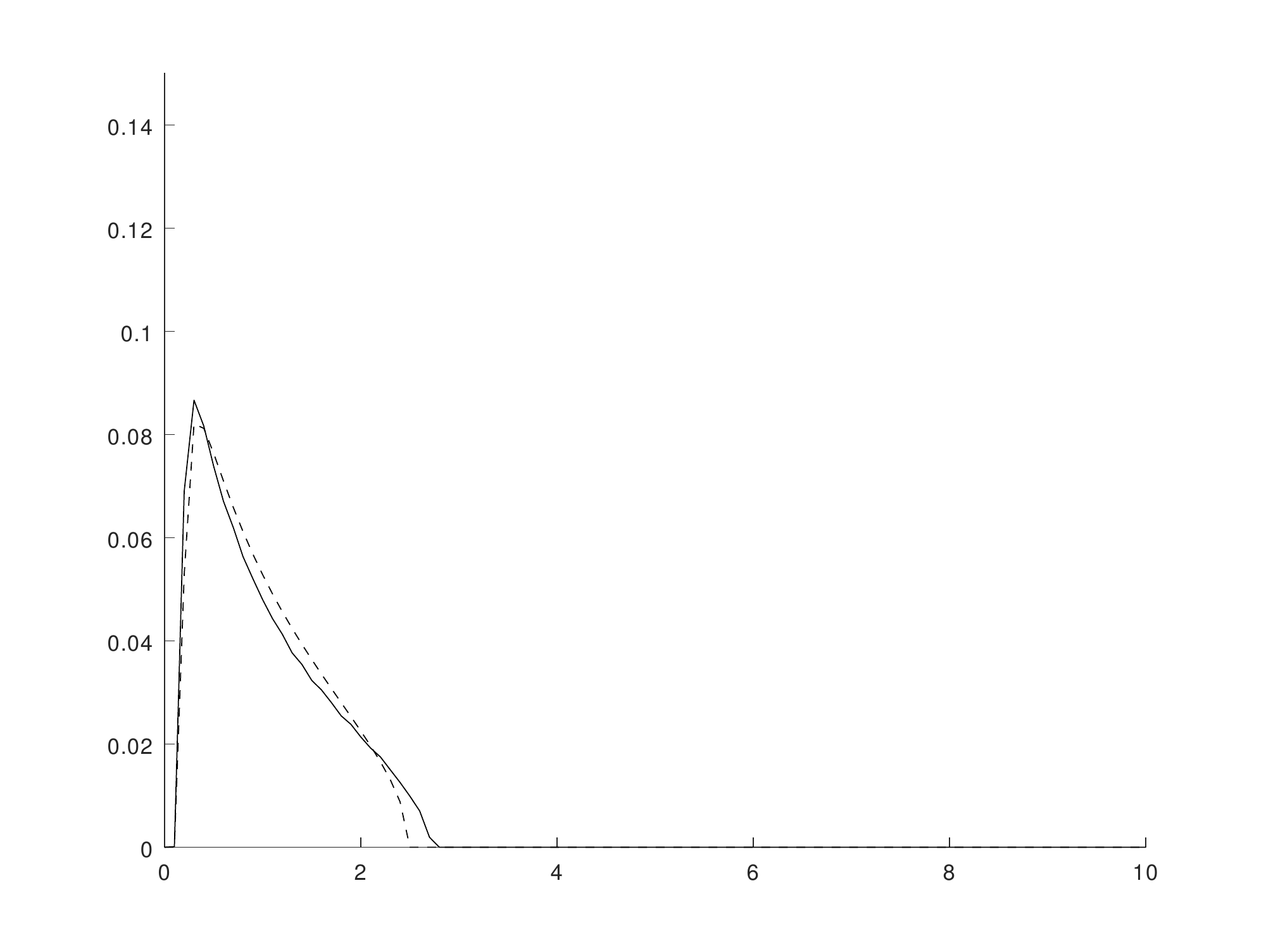} 
\hspace{1.6cm} (b) $H=0.2$
\end{center}
\end{minipage}
 &
 \begin{minipage}{0.33\hsize}
\begin{center}
 \includegraphics[width=5.5cm]{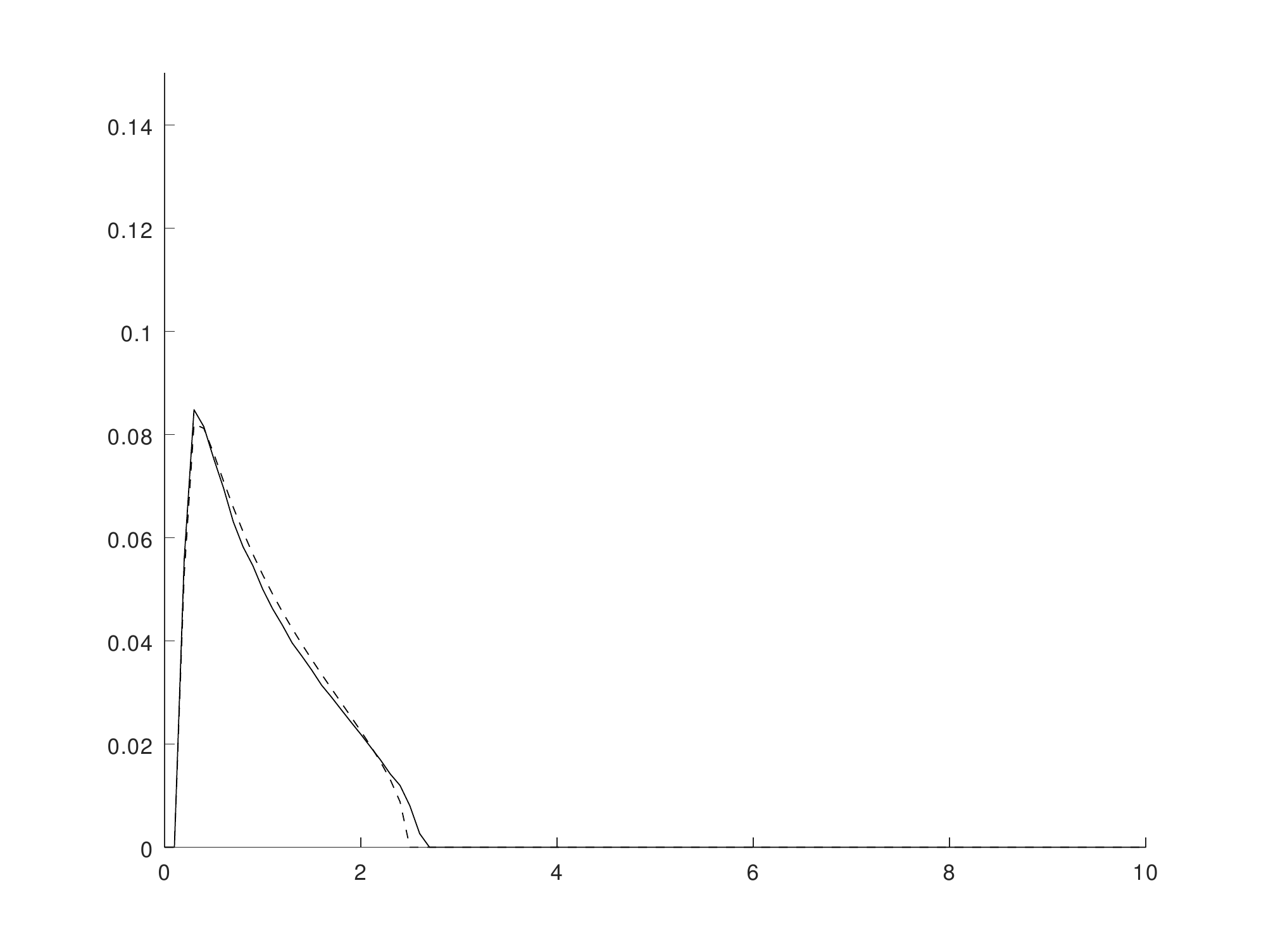} 
\hspace{1.6cm} (c) $H=0.3$
\end{center}
\end{minipage} 
\\
\begin{minipage}{0.33\hsize}
\begin{center}
 \includegraphics[width=5.5cm]{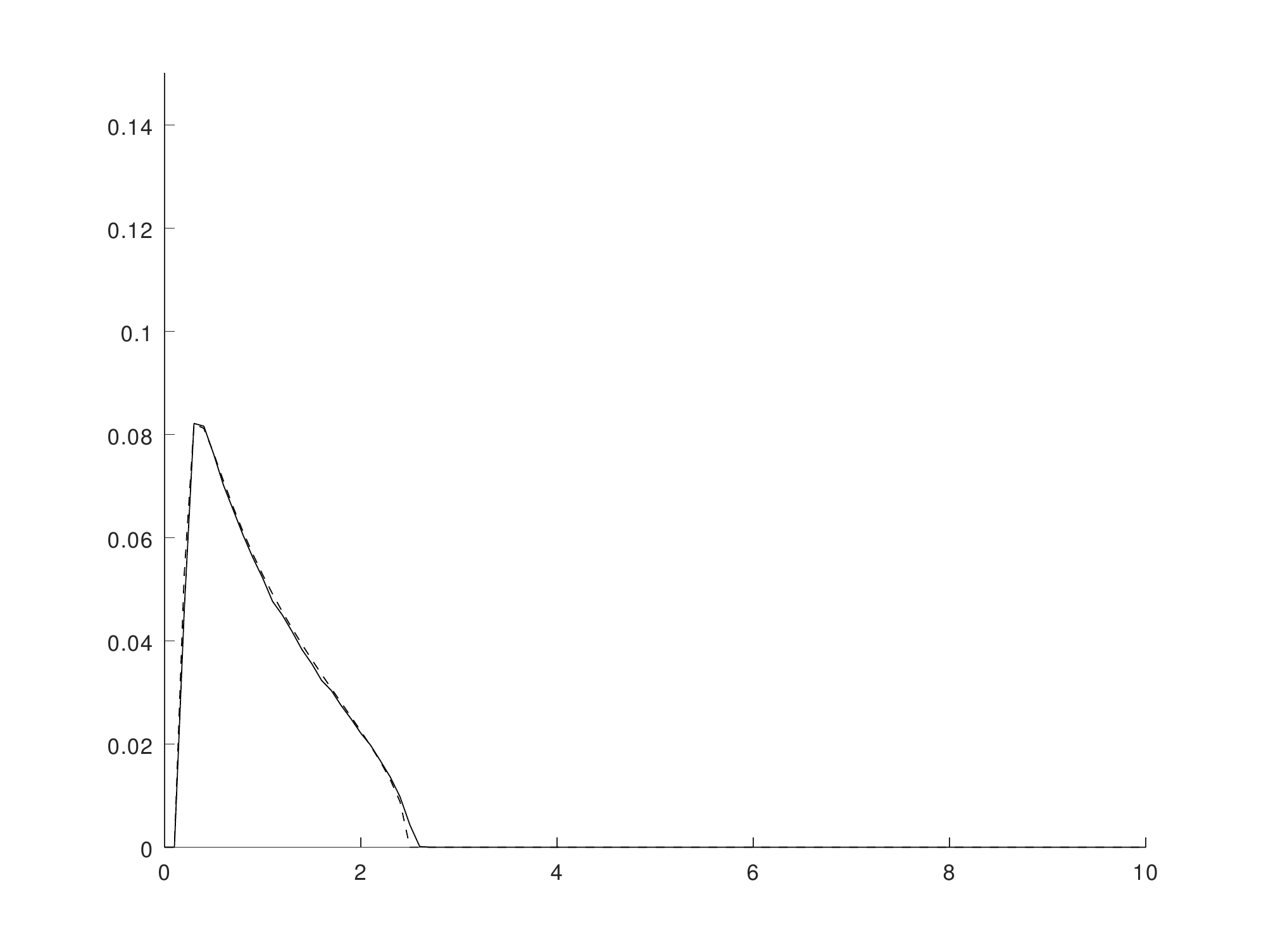} 
\hspace{1.6cm} (d) $H=0.4$
\end{center}
\end{minipage}
 &
\begin{minipage}{0.33\hsize}
\begin{center}
 \includegraphics[width=5.5cm]{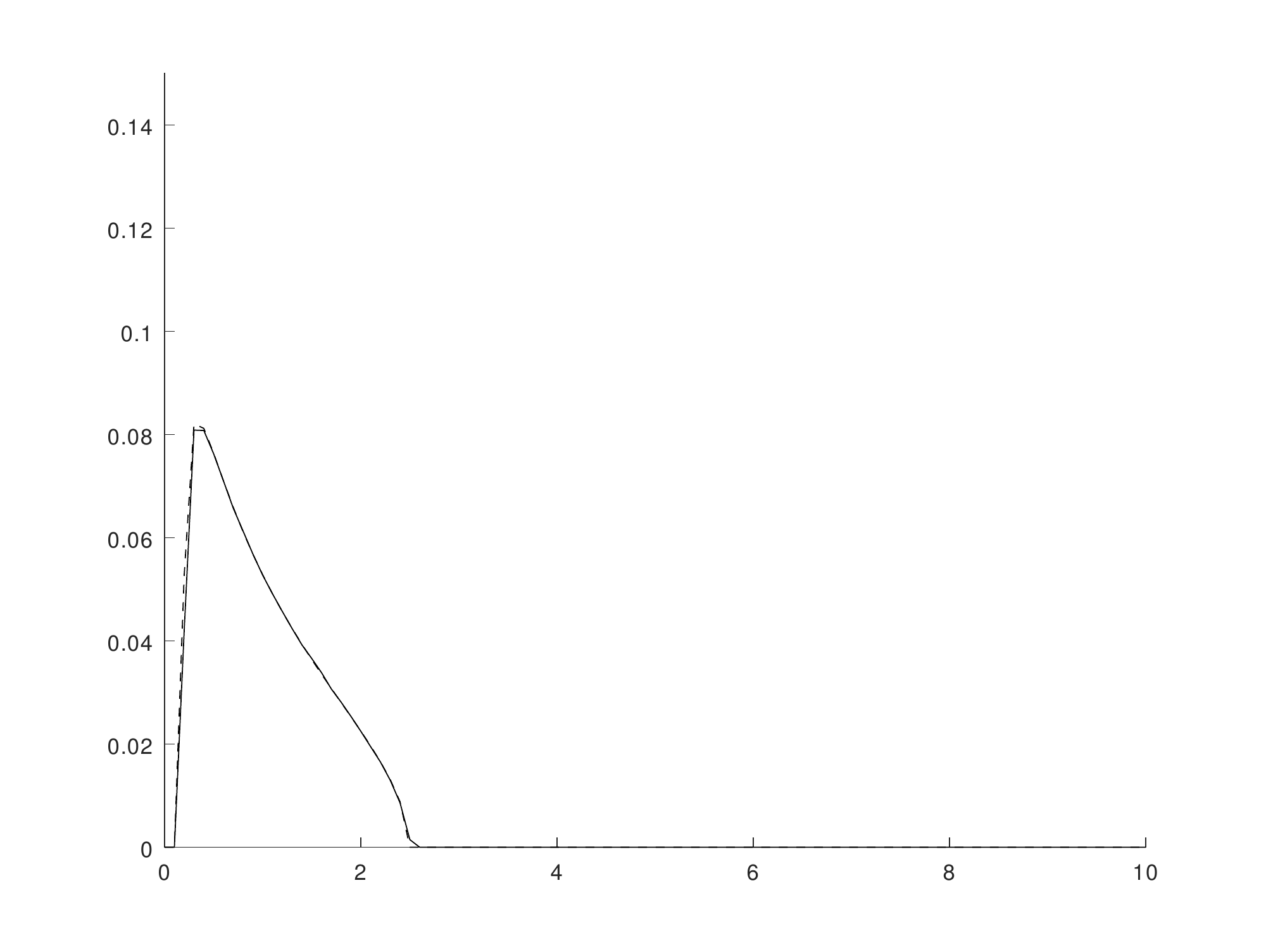} 
\hspace{1.6cm} (e) $H=0.5$
\end{center}
\end{minipage}
 &
 \begin{minipage}{0.33\hsize}
\begin{center}
 \includegraphics[width=5.5cm]{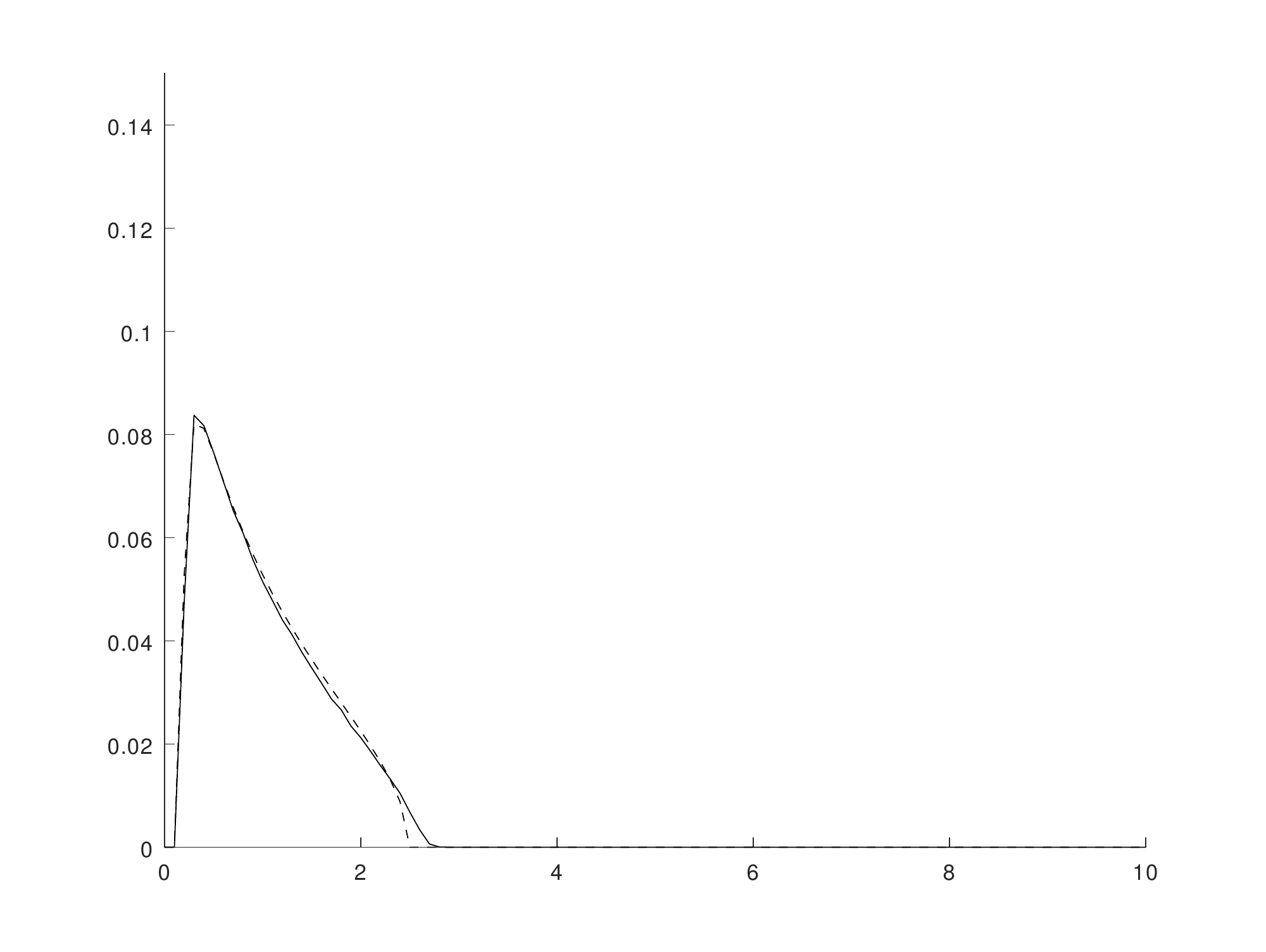} 
\hspace{1.6cm} (f) $H=0.6$
\end{center}
\end{minipage} 
\\
\begin{minipage}{0.33\hsize}
\begin{center}
 \includegraphics[width=5.5cm]{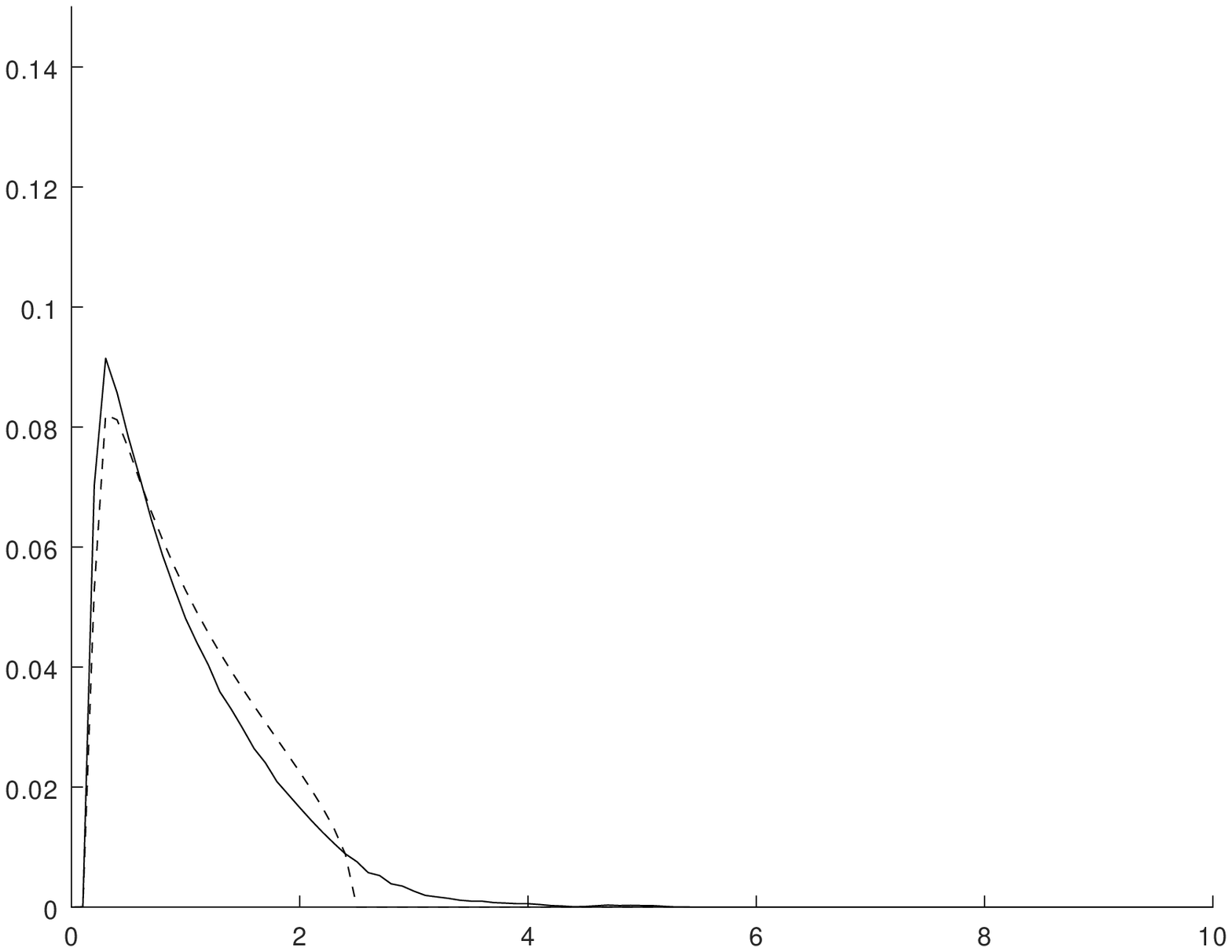} 
\hspace{1.6cm} (g) $H=0.7$
\end{center}
\end{minipage}
 &
\begin{minipage}{0.33\hsize}
\begin{center}
 \includegraphics[width=5.5cm]{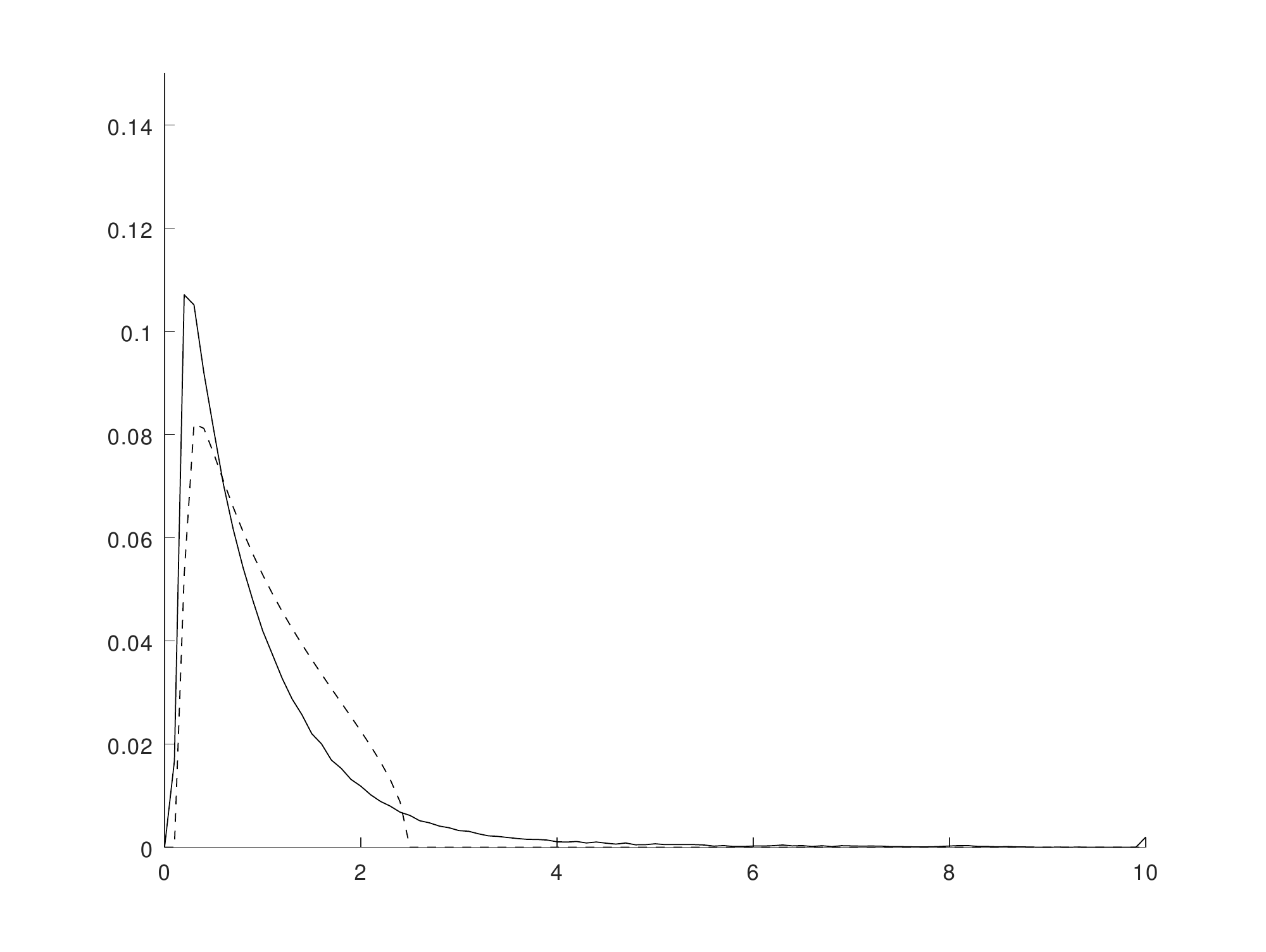} 
\hspace{1.6cm} (h) $H=0.8$
\end{center}
\end{minipage}
 &
 \begin{minipage}{0.33\hsize}
\begin{center}
 \includegraphics[width=5.5cm]{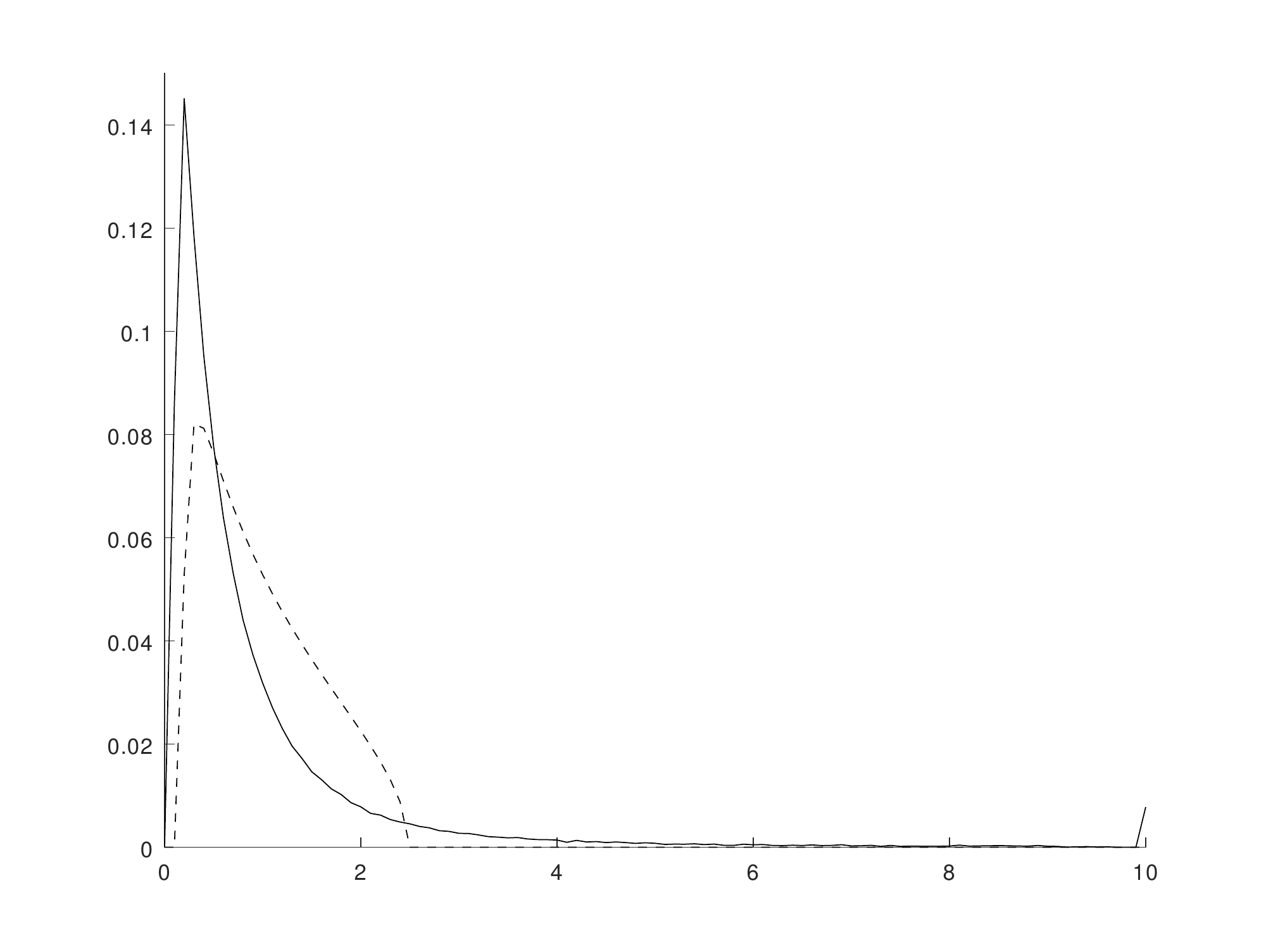} 
\hspace{1.6cm} (i) $H=0.9$
\end{center}
\end{minipage}  \\
\end{tabular}  
\caption{Plots of the histogram of  the deformed MPD for $Q=3$   for  fBm :
(a)$H=0.1$, (b)$H=0.2$, (c)$H=0.3$,
(d)$H=0.4$, (e)$H=0.5$, (f)$H=0.6$,
(g)$H=0.7$, (h)$H=0.8$, (i)$H=0.9$.
The horizontal axis represents  the eigenvalues and the vertical axis 
is the frequency.
The real line is the distribution with the correlation and the dotted line is the MPD. We can confirm the fat tail distribution for large and small $H$.
}
\label{MPD2}
\end{figure}

First, we confirm  the distributions of the 
deformed MPD for  the fBm case.
We  calculated 100 times for  each $H$ and 
created a histogram of the eigenvalues.
The conclusions are shown in Fig.\ref{MPD2}.
We can confirm the difference from the  MPD in smaller $H$ 
and larger $H$. 
On the other hand  for around $H=0.5$, the distributions have   almost the  same shape as  the MPD, because $H=1/2$ is the Brownian motion.
Above $H_c=3/4$, there is  a  phase transition and
we can observe  large eigenvalues.
In fact we can observe the  peak at  $10$  in $H=0.8$ and $0.9$  in Fig.\ref{MPD2} (h) and (i) which corresponds to the large eigenvalues above 10.

\subsubsection{ii. Convergence of the moments}
\begin{figure}[htbp]
\begin{center}
\begin{tabular}{cc}
 \begin{minipage}{0.5\hsize}
\begin{center}
\includegraphics[width=8cm]{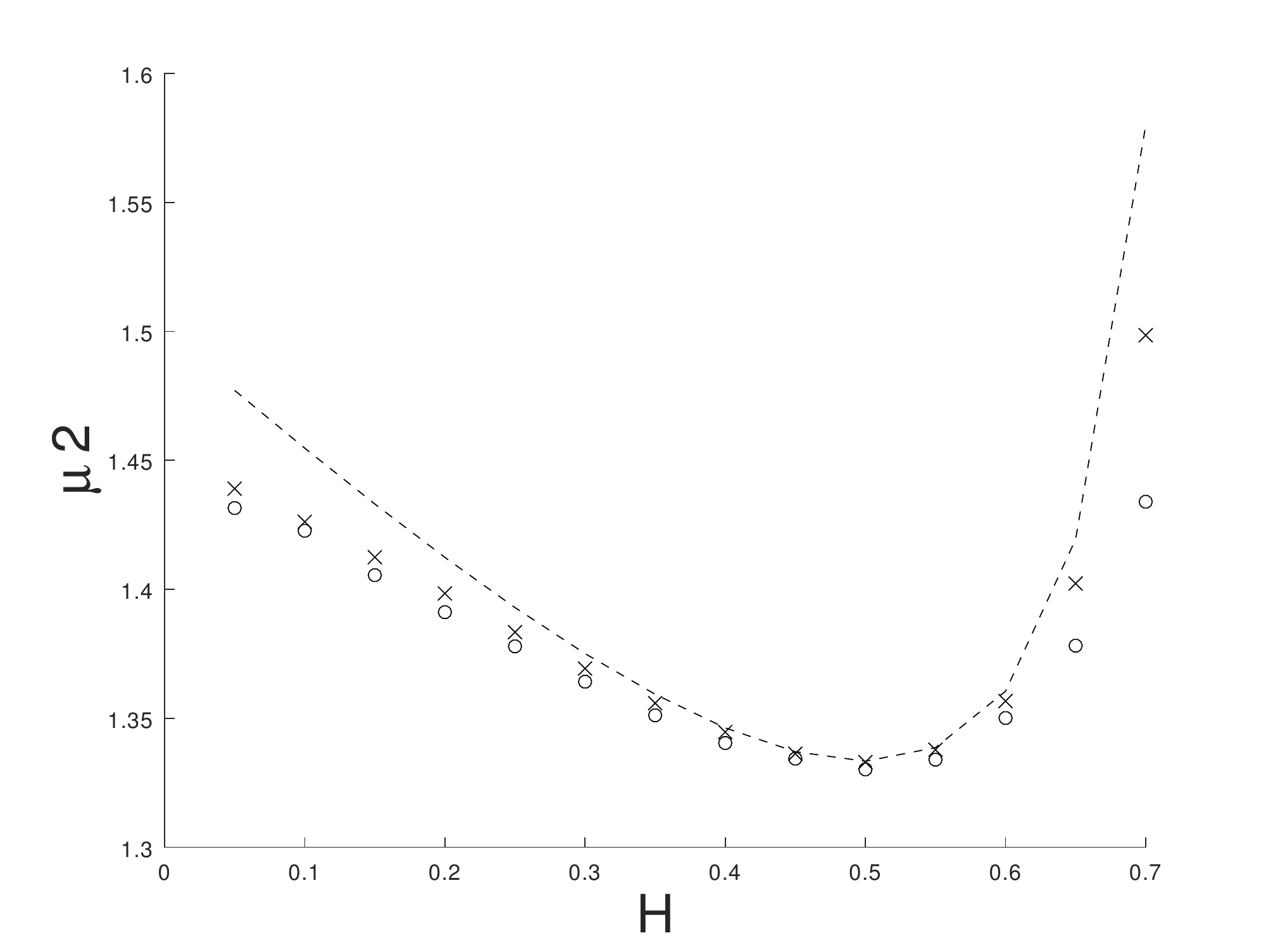}
\hspace{1.6cm} (a)
\end{center}
\end{minipage}
& 
\begin{minipage}{0.5\hsize}
\begin{center}
\includegraphics[width=8cm]{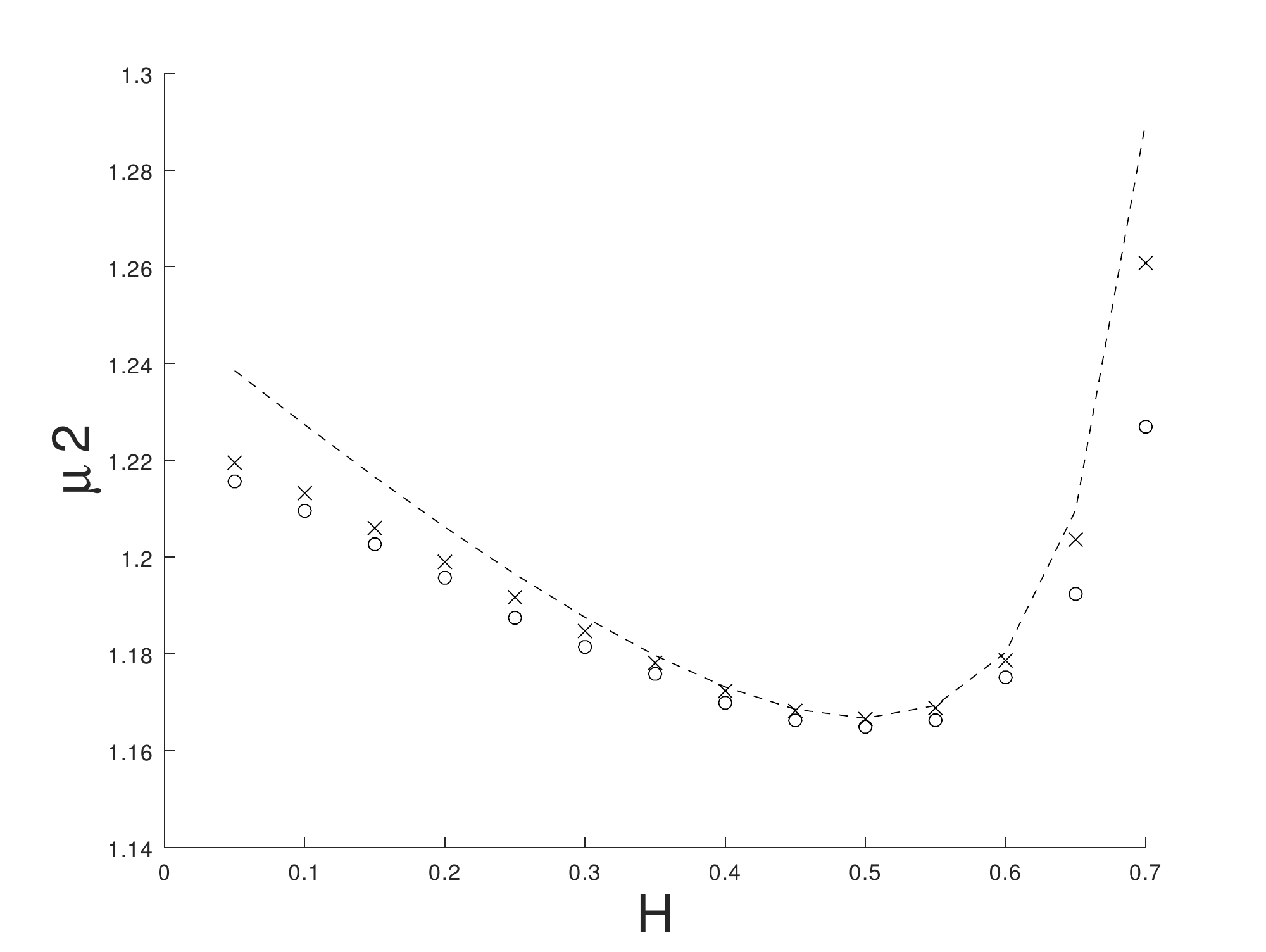}
\hspace{1.6cm} (b) 
\end{center}
\end{minipage}
\end{tabular}
\end{center}
\caption{
The figure (a)  shows  $\mu_2$ in $Q=3$ and the figure (b) shows $Q=6$ in the case of fBm.
These are the comparisons between the theoretical moments and 
$N=512, 64$  numerical simulations.
The horizontal axis is $H$ and the vertical axis is the second moment.
"x", "o" indicate $N=512$, $64$ and the dotted lines show the theoretical moments. 
}
\label{concent2}
\end{figure}

Next,  we confirm the convergence of the distributions for  the  fBm case.
In Fig.\ref{concent2} we show  $\mu_2$ in $Q=3,6$ below $H_c=3/4$,  which is the transition point.
Above $H_c=3/4$, the second moment diverges.
The horizontal axis is $H$ and the vertical axis represents the second  moment.
As the matrix size increases,  the moments converge  to
the theoretical moments.
Appendix C  presents the 
conclusions of the numerical simulations in detail.

\section{V. Phase transition of the deformed   Marchenko-Pastur distribution }
In this section, we study the phase transition  in the case of  power decay.
We discuss  the convergence of the scaled  largest  eigenvalue  which is the order parameter  of  the phase transition in subsection A  and  apply  finite scaling analysis to confirm the critical exponent in  subsection B. 

\subsection{A. The scaled largest eigenvalue and phase transition}
\begin{figure}[htbp]
\begin{center}
\begin{tabular}{cc}
 \begin{minipage}{0.5\hsize}
\begin{center}
\includegraphics[width=8cm]{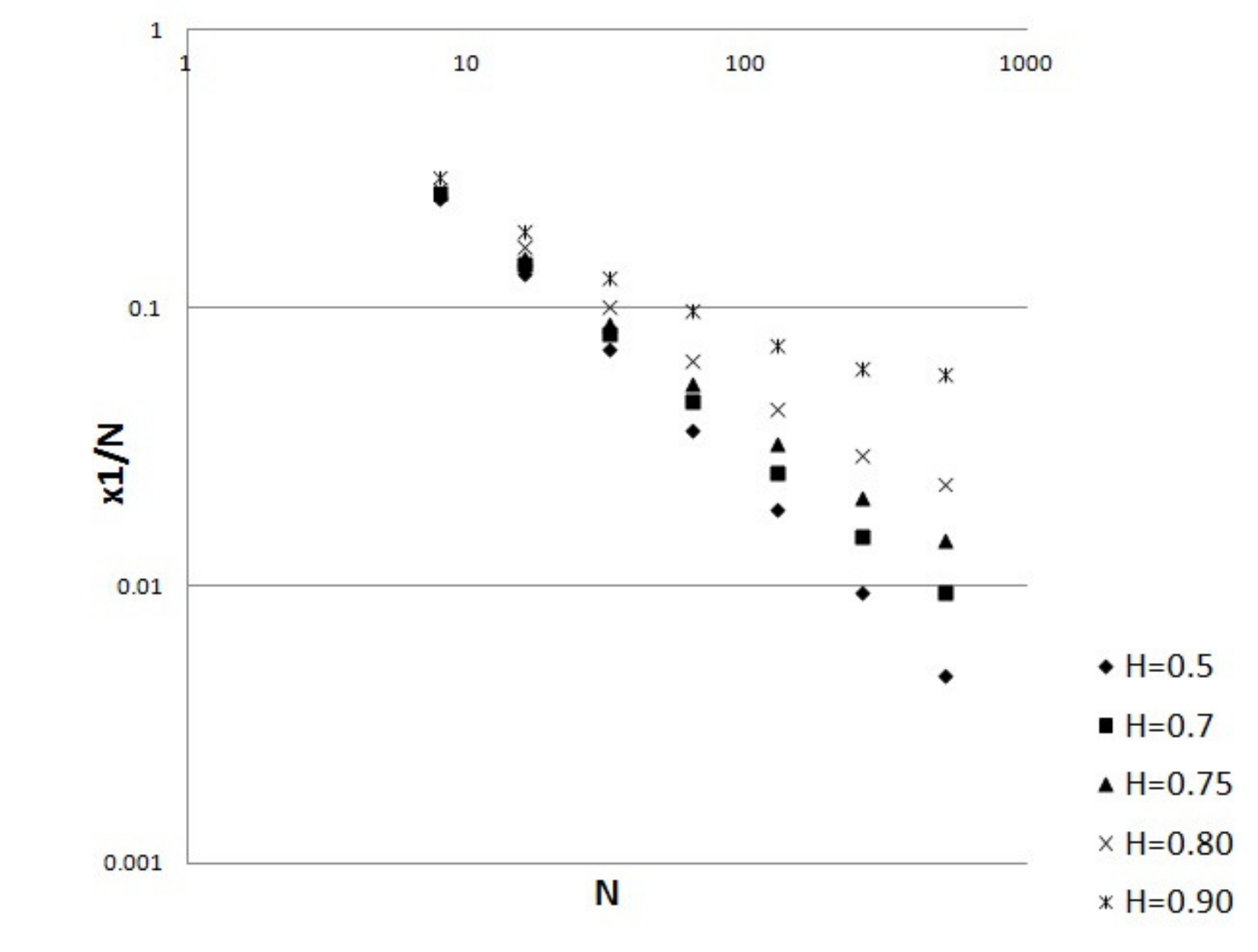}
\hspace{1.6cm} (a)
\end{center}
\end{minipage}
& 
\begin{minipage}{0.5\hsize}
\begin{center}
\includegraphics[width=8cm]{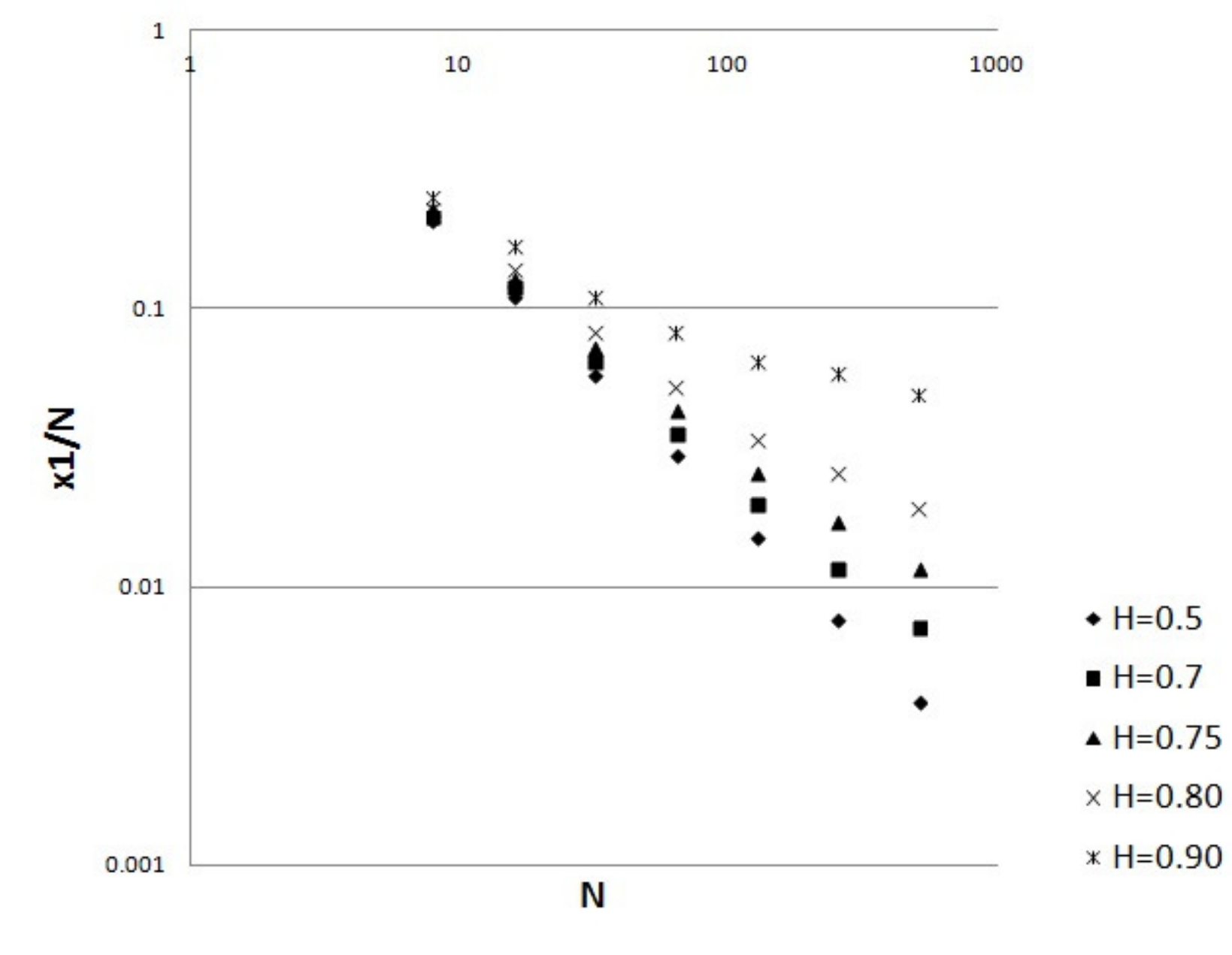}
\hspace{1.6cm} (b) 
\end{center}
\end{minipage}
\end{tabular}
\end{center}
\caption{
Figures (a)  and (b) show    the scaled  largest eigenvalue, $x_1/N$, when $Q=3$ and $Q=6$   for fBm.
The horizontal axis is a  log  of the matrix size $N$,  and the vertical axis is  a log  of the  scaled largest eigenvalue $x_1/N$.
We plot the cases  $H=0.5, 0.7, 0.75, 0.8, 0.9$.
When $H<Hc=3/4$, the scaled largest eigenvalue  is on a straight line and   converges to $0$.
When $H\geq Hc=3/4$,  the scaled largest eigenvalue is  not on a  straight line.}
\label{concent4}
\end{figure}

In this subsection, we confirm the convergence of the scaled  largest eigenvalue, $0<x_1/N\leq 1$ in the case of fBm.
In Fig.\ref{concent4} we show the 
 scaled largest eigenvalues   when $Q=3$ and $Q=6$.
 The horizontal axis is   a log plot of  matrix size $N$ and the vertical axis is  a log plot of the  scaled largest eigenvalue $x_1/N$.
When $H<Hc=3/4$, the scaled largest eigenvalue  is on a straight line and   converges to $0$, because the largest eigenvalue is finite.
When $H\geq Hc=3/4$,  the scaled largest eigenvalue is  not on a straight line.
The dominance  of the  largest eigenvalue  changes at  $H_c$ and increases above $H_c$.
Therefore,  the scaled    largest eigenvalue is the order parameter of the phase transition. 

\subsection{B. Finite size scaling}

In this subsection, we consider  finite size scaling.
We introduce the scaling function to  the scaled  largest eigenvalue,  $x_1 /N$,  which  we discussed in the previous subsection,
\begin{equation}
    \frac{x_1}{N}=N^{\gamma/\nu}f(N^{1/\nu}t), 
    \label{sc}
\end{equation}
where
$t=(H-H_c)/H_c$ and $f(x)$ is a scaling function.
It is the hypothesis that  the data of several $N$ are on  the curve, Eq.(\ref{sc}).  
We assume that  the scaling function is $f(x)\sim 0$  in the limit $x\rightarrow -\infty$ and $f(x)\sim O(x)$   in the limit $x \rightarrow \infty$.
In these hypotheses, we can obtain the following  relation,
\begin{eqnarray}
m\equiv \frac{x_1}{N}&\propto& t=(H-H_c)/H_c \hspace{2cm} H-H_c>>N^{-1/\nu}
\nonumber  \\ 
&\propto& 0. \hspace{5cm} H_c-H>>N^{-1/\nu}
\label{sc2}
\end{eqnarray}
We confirmed  convergence to $0$ in the case, $H<H_C$ in the previous subsection.

\begin{figure}[htbp]
\begin{center}
\begin{tabular}{cc}
 \begin{minipage}{0.5\hsize}
\begin{center}
\includegraphics[width=8cm]{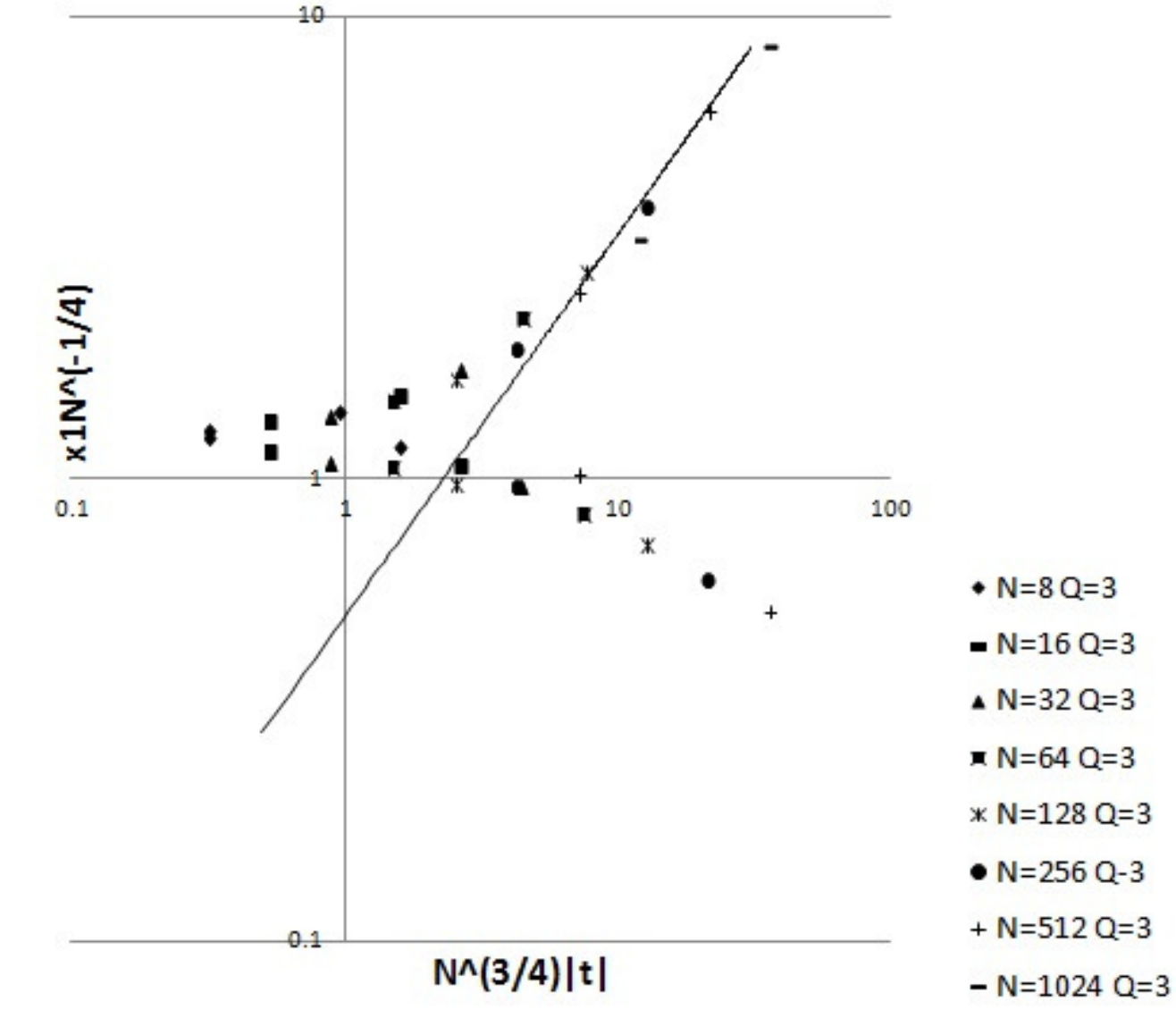}
\hspace{1.6cm} (a)
\end{center}
\end{minipage}
& 
\begin{minipage}{0.5\hsize}
\begin{center}
\includegraphics[width=8cm]{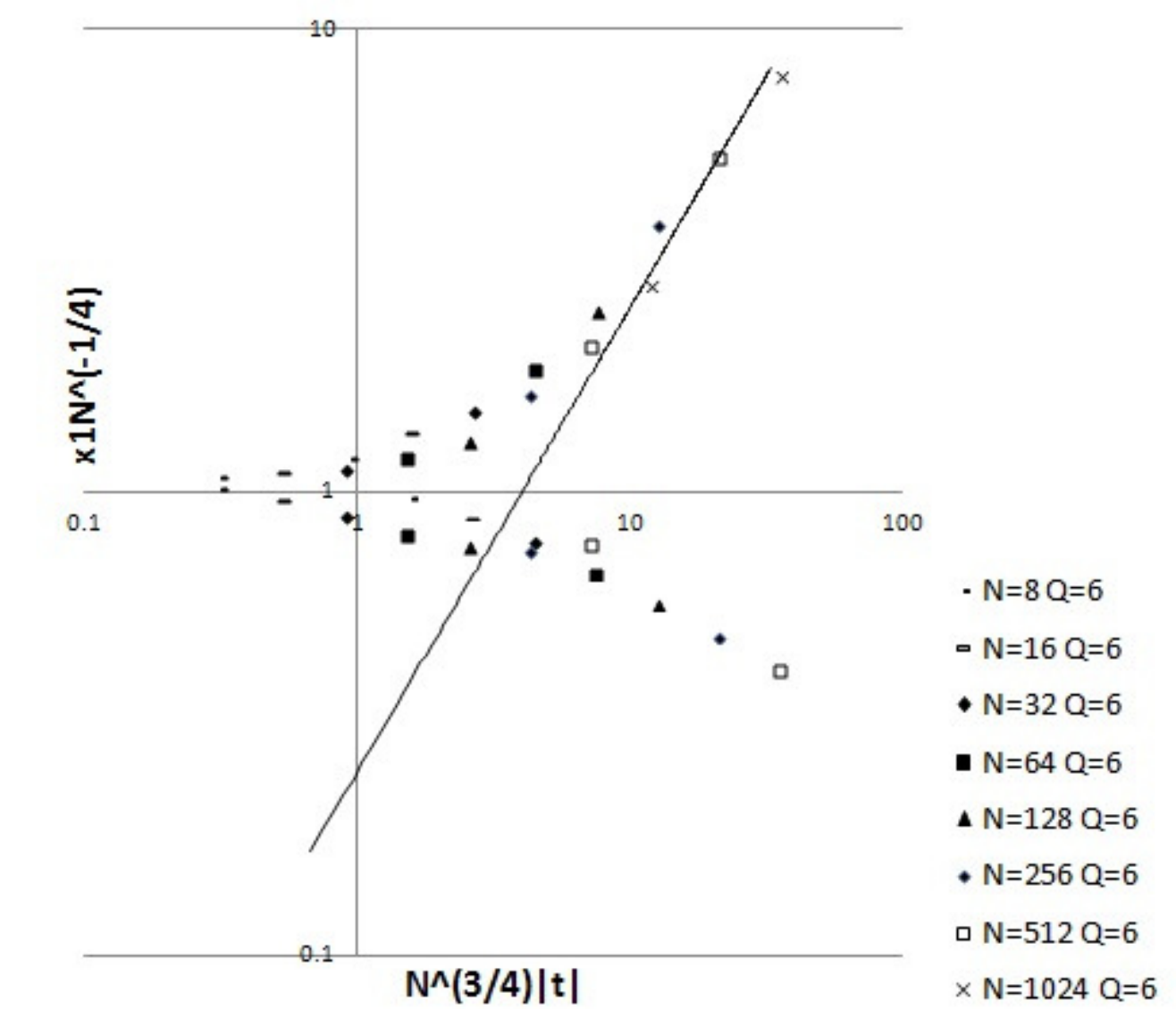}
\hspace{1.6cm} (b) 
\end{center}
\end{minipage}
\end{tabular}
\end{center}
\caption{
Figures (a)  and (b) show    the finite size scaling plot when $Q=3$ and $Q=6$ in the case of power decay.
The horizontal axis is  a log  of $N^{3/4}|t|=N^{3/4}|H-H_c|/H_c$, RHS of  Eq.(\ref{sc}) and 
the vertical axis is a log  of $x_1/N^{1/4}$, LHS of Eq.(\ref{sc}).
Here we use $H_c=3/4$.
We can  confirm that  $N=8,16,32, 64,128, 256, 512, 1024$ are on a curve  near the critical point and the asymptotic line is $x$ which corresponds to the scaling function.
The solid line is the asymptotic line  $t$ and the trend is $1$  which is consistent with Eq.(\ref{sc2}).
}
\label{scaling}
\end{figure}

We confirm    this scaling  hypothesis in Fig \ref{scaling}, where we set  $\gamma=-1$ and $\nu=4/3$.
We estimate  these parameters by setting  that the data of  several $N$ are on the curve.
The slope of the asymptotic line   is $1$ in large $t$,
 and it is consistent with   Eq.(\ref{sc}) and  Eq.(\ref{sc2}).
 
Hence, we can estimate the critical exponent,
\begin{equation}
m\propto  t^{\beta} \hspace{2cm}{\rm at} \hspace{2cm}H\sim H_c
\end{equation} where $\beta=1$ and
\begin{equation}
    \xi_{\infty}\propto t^{-\nu}=t^{ -4/3} \hspace{0.8cm}{\rm at} \hspace{2cm}H\sim H_c,
\end{equation}
where $\xi_{\infty}$ is the correlation length.

\section{VI. Concluding Remarks}
We considered the   time series  with the correlation and its Wishart matrix.
When the variables are independent,   the 
eigenvalue  distribution of the Wishart matrix  converges to the Marchenko-Pastur distribution (MPD).
When there is  a correlation, the eigenvalue distribution
converges to the deformed MPD.
The deformed MPD has a  long tail and a  high peak for large temporal correlations.
We calculated   some moments of the distribution  for the exponential and power decay cases and 
 discussed the convergence of this distribution.
We  have shown  that the mean of the distribution does not depend on the correlation and that  the second moment increases as the temporal correlation increases.
In particular when the temporal correlation is power decay, we can confirm a phenomenon  such as  phase transition from the finite second moment to the infinite second moment.
In other words,  it is the transition  between the
finite largest eigenvalue and the infinite largest eigenvalue.
If $\gamma>1/2$ which is the power index, the second moment of the distribution  and the  largest eigenvalue are finite.
On the other hand,   when $\gamma\leq 1/2$, the  second moment and the largest  eigenvalue are  infinite.
The largest eigenvalues, the explicit formula of  the deformed MPD  and the  closed form  of the higher moments,  especially for the power decay case,  are  future problems.

The other future problem is the study of 
phase transition for the power decay case.
We discussed  the convergence of the scaled  largest  eigenvalue  which is the order parameter  of  the phase transition and  applied  finite scaling analysis and  estimate the critical exponent.
The theoretical confirmations are  the future problems.

\def\thesection{Appendix \Alph{section}}

\section{Appendix A  the Marchenko-Pastur distribution  and its moments}
In this Appendix, 
we review the moments of the Marchenko-Pastur distribution (MPD).
The MPD is 
\[
P(\lambda)=\frac{Q}{2\pi \lambda}\sqrt{(\lambda_+-\lambda)(\lambda-\lambda_-)},
\]
where 
\[\lambda_{\pm}=1+\frac{1}{Q}\pm2\sqrt{\frac{1}{Q}},
\]
and
\[Q=\frac{L}{N},
\]
in the limit $N, L\rightarrow \infty$ with  constant $Q$.

We can obtain the moment of the MPD
\begin{eqnarray}
\mu_k&=&E(\lambda^k)=\int_{\lambda_-}^{\lambda_+}\lambda^k P(\lambda)d\lambda\nonumber \\
&=&
\frac{2}{\pi}\int_{-1}^{1}
[2\sqrt{\frac{1}{Q}}x+(1+\frac{1}{Q})]^{k-1}
\sqrt{1-x^2}dx,
\end{eqnarray}
where 
\[x=\frac{\lambda-(1+\frac{1}{Q})}{2\sqrt{\frac{1}{Q}}}.
\]
Here we divided the cases when $k=2m+1$  and $k=2m+2$,$m=1,2,\cdots$.

In the case $k=2m+1$, we expand the first term by $x$    and 
the odd powers of $x$  become 0,
\begin{eqnarray}
\mu_k&=&
\sum_{i=0}^m
\left(
    \begin{array}{c}
      2m  \\
      2i
    \end{array}
  \right)
  C_i
  (1+\frac{1}{Q})^{2m-2i}\frac{1}{Q^i}
  \nonumber \\
  &=&
\sum_{j=1}^{k}   
\sum_{i=0}^m
\left(
    \begin{array}{c}
      2m  \\
      2i
    \end{array}
  \right)
    \left(
    \begin{array}{c}
      2m-2i  \\
      j-1-i
    \end{array}
  \right)
  C_i
  \frac{1}{Q^{j-1}}
  \nonumber \\
  &=&
  \sum_{j=1}^k
  \frac{1}{k}
  \left(
    \begin{array}{c}
      k  \\
      j
    \end{array}
  \right)
 \left(
    \begin{array}{c}
      k  \\
      j-1
    \end{array}
  \right)
  \frac{1}{Q^{j-1}}
  \\
  \label{mk1}
  &=&\sum_{j=1}^k
  N(k,j)\frac{1}{Q^{j-1}}
  =
  \mathcal{N}_k(\frac{1}{Q})
  \label{np1}
\end{eqnarray}
$N(k,j)$ is the Narayana number and $\mathcal{N}_k(x)$ is the  Narayana polynomial.
Then, the momentum of the MPD is the  Narayana polynomial.

In the first first equal  we use the relation
  \[
  \int^1_{-1}x^{2i}\sqrt{1-x^2}dx=\frac{\pi}{2^{2i+1}}C_i,
  \]
  and  
  $C_i$ is the $i$ the Catalan number 
\begin{eqnarray}
C_i=\frac{1}{i+1}
  \left(
    \begin{array}{c}
      2i \\
      i
    \end{array}
  \right).
\end{eqnarray}
In the second equal  we use $j=i+l+1$ instead of $l$.
In the third  equal  we use  the  following identity,  
\begin{eqnarray}
\frac{1}{k}
  \left(
    \begin{array}{c}
      k  \\
      j
    \end{array}
  \right)
 \left(
    \begin{array}{c}
      k  \\
      j-1
    \end{array}
  \right)
  =\sum_{i=0}^{\lfloor (k-1)/2\rfloor}
   \left(
    \begin{array}{c}
      k-1 \\
      2i
    \end{array}
  \right)
 \left(
    \begin{array}{c}
      k -2i-1 \\
      j-1-i
    \end{array}
  \right) 
  C_i.
  \end{eqnarray}

When $k=2m+2$ we can obtain in the same way
\begin{eqnarray}
\mu_k&=&\sum_{i=0}^m
\left(
    \begin{array}{c}
      2m+1  \\
      2i
    \end{array}
  \right)
  C_i
  (1+\frac{1}{Q})^{2m+1-2i}
  \frac{1}{Q^i}
  \nonumber \\
  &=&\sum_{j=1}^k
  N(n,j)\frac{1}{Q^{j-1}}=\mathcal{N}_k(\frac{1}{Q}).
  \label{np2}
\end{eqnarray}

In summary we can obtain the moment $\mu_k$ for the MPD  as  Narayana  polynomial  \cite{Yang},
\begin{eqnarray}
\mu_1&=&1
\nonumber \\
\mu_2&=&1+\frac{1}{Q}
\nonumber \\
\mu_3&=&1+\frac{3}{Q}+\frac{1}{Q^2}
\nonumber \\
\mu_4&=&
1+\frac{6}{Q}+\frac{6}{Q^2}+\frac{1}{Q^3}
\nonumber \\
\mu_5&=&1+\frac{10}{Q}+\frac{20}{Q^2}+\frac{10}{Q^3}+
\frac{1}{Q^4}
\nonumber \\
\mu_6&=&1+\frac{15}{Q}+\frac{50}{Q^2}+\frac{50}{Q^3}+\frac{15}{Q^4}+\frac{1}{Q^5}.
\nonumber \\
\cdots
\end{eqnarray}


\section{Appendix B Fractional Brownian motion}
In this Appendix, we review the  fractional Brownian motion (fBm).
$H\in (0,1)$ is the parameter of the fBm.
When the Gauss process $B^H=\{ B_t^H\}_{t\geq1}$ with $E(B^H)=0$ satisfies following condition, the process is 
 the fBm,
\begin{equation}
\mbox{Cov}(B_S^H, B_t^H))= \frac{1}{2}
(|t|^{2H}+|s|^{2H}-|t-s|^{2H}),
\end{equation}
in any $t$ and $s$.
$H$ is the Hurst index.
In the  case $H=1/2$, the process is    Brownian motion.
In the case $H\neq1/2$, the process has a power decay temporal correlation with $\gamma\sim 2-2H$ in  $t>>1$.
Here $\gamma$ is the power index.
Hence, when  $H>1/2$, the process has  long memory and
$H<1/2$  the process has short memory.

\section{Appendix C  Convergence to the theoretical values of the deformed MPD}
In this Appendix  we compare the theory and  numerical simulations.
We show the cases $Q=3$ and $Q=6$ of exponential decay temporal correlation in Tables.\ref{Q=3} and \ref{Q=6}, respectively.
In Tables.\ref{FBM2} and  \ref{FBM3} we show the cases of the  fBm  in $Q=3,6$.
At $H=3/4$ $\mu_2$ becomes infinite.

\begin{table}[tbh]
\caption{Comparison  of the moments between the simulations and theory  for $Q=3$}
\begin{center}
\begin{tabular}{|l|l|c|c|c|c|c|c|c|c|c|}
\multicolumn{4}{c}{}\\ \hline
$r$&0& 0.1 &0.2&0.3&0.4&0.5&0.6&0.7&0.8&0.9\\ 
 \hline \hline
$N=512, \mu_2$&
1.3329	&	1.3396	&	1.3605	&	1.3984	&	1.4591	&	1.5537	&	1.7053	&	1.9681	&	2.5047	&	4.1215
\\ \hline
$N=64, \mu_2$&
1.3295	&	1.3364	&	1.3565	&	1.3934	&	1.4508	&	1.5405	&	1.6836	&	1.9297	&	2.4130	&	3.7446
\\ \hline
Theory $\mu_2$ &
1.3333 	&	1.3401 	&	1.3611 	&	1.3993 	&	1.4603 	&	1.5556 	&	1.7083 	&	1.9739 	&	2.5185 
& 4.1754 

\\ \hline \hline
$N=512, \mu_3$ &
2.1092	&	2.1362	&	2.221	&	2.3775	&	2.6378	&	3.0651	&	3.8058	&	5.2536	&	8.8539	&	24.9533
\\ \hline 
$N=64, \mu_3$&
2.0950	&	2.1228	&	2.2034	&	2.3549	&	2.5986	&	
2.9993 &	3.6891	&	5.0216	&	8.1719	&   20.5884
\\ \hline
Theory $\mu_3$ &
2.11111 	&	2.13812 	&	2.22338 	&	2.38137 	&	2.64324 	&	3.07407 	&	3.82205 	&	5.28861 	&	8.95885 	&	25.59588 
\\ \hline \hline
$N=512, \mu_4$ &
3.698	&	3.7792	&	4.0378	&	4.5286	&	5.3836	&	6.8831	&	9.746	&	16.1855	&	36.184	&	173.524
\\ \hline 
$N=64, \mu_4$&
3.6555	&	3.7394	&	3.9837	&	4.4554	&	5.2481	&	
6.6385&	9.2643	&	15.0777	&	32.0347	&	130.7156
\\ \hline
Theory $\mu_4$ &
3.70370 	&	3.80732 	&	4.13587 	&	4.75033 	&	5.78594 	&	7.53909 	&	10.73691 	&	17.59557 	&	38.10822 	&	180.77500 
\\ \hline 
\end{tabular}
\label{Q=3}
\end{center}
\end{table}

\begin{table}[tbh]
\caption{Comparison of the moments between the simulations and theory  for $Q=6$}
\begin{center}
\begin{tabular}{|l|l|c|c|c|c|c|c|c|c|c|}
\multicolumn{4}{c}{}\\ \hline
$r$&0& 0.1 &0.2&0.3&0.4&0.5&0.6&0.7&0.8&0.9\\ 
 \hline \hline
$N=512, \mu_2$&1.1671	&	1.1698	&	1.1802	&	1.1992	&	1.2295	&	1.2771	&	1.3532	&	1.4853	&	1.7554	&	2.5731	 \\ \hline
$N=64, \mu_2$&1.1647	&	1.1675	&	1.1782	&	1.19660	&	1.2260	&	1.2718	&	1.3455	&	1.4716	&	1.7247	&	2.4649	 \\ \hline
Theory $\mu_2$ &
1.16667 	&	1.17003 	&	1.18056 	&	1.19963 	&	1.23016 	&	1.27778 	&	1.35417 	&	1.48693 	&	1.75926 	&	2.58772 
\\ \hline \hline
$N=512, \mu_3$ &
1.5294	&	1.5386	&	1.5753	&	1.6432	&	1.7536	&	1.9322	&	2.2326	&	2.7954	&	4.1076	&	9.418
\\ \hline 
$N=64, \mu_3$&
1.5209	&	1.5308	&	1.5681	&	1.6337	&	1.7400	&	1.9112	&	2.1998	&	2.7308&	3.9402	&	8.5995	 \\ \hline
Theory $\mu_3$ &
1.52778 	&	1.53958 	&	1.57668 	&	1.64479 	&	1.75605 	&	1.93519 	&	2.23676 	&	2.80254 	&	4.12860 	&	9.53055 
\\ \hline \hline
$N=512, \mu_4$ &
2.1756	&	2.1986	&	2.2924	&	2.4685	&	2.7634	&	3.2614	&	4.1568	&	6.0069	&	11.0852	&	39.8625
\\ \hline 
$N=64, \mu_4$&
2.1540	&	2.7193	&	2.2740	&	2.4436	&	2.7257	&	
3.2013&	4.0535	&	5.7781	&	10.3701	&	34.7996	 \\ \hline
Theory $\mu_4$ &
2.17129	&	2.21626	&	2.35784 	&	2.61857 	&	3.04684 	&	3.74331 	&	4.93717 	&	7.25797 	&	13.15499 	&	43.86338 
\\ \hline 
\end{tabular}
\label{Q=6}
\end{center}
\end{table}

\begin{table}[tbh]
\caption{Comparison  of the moments between the simulations and theory  for fBm in $Q=3$}
\begin{center}
\begin{tabular}{|l|c|c|c|c|c|c|c|c|c|}
\multicolumn{4}{c}{}\\ \hline
$H$& 0.1 &0.2&0.3&0.4&0.5&0.6&0.7&0.8&0.9\\ 
 \hline \hline
$N=512, \mu_2$&
1.4261	&	1.3982	&	1.369	&	1.3448	&	1.3331	&	1.3568	&	1.4991& 2.1279&4.7319\\
\hline 
$N=64, \mu_2$&
1.4159	&	1.3894	&	1.3648	&	1.3407	&	1.3289	&	1.3485	&	1.4338& 1.6687&2.2148\\
\hline 
Theory&
1.454	&	1.412	&	1.375	&	1.346	&	1.333	&	1.360	&	1.579& N.A.& N.A.
\\
\hline
\end{tabular}
\label{FBM2}
\end{center}
\end{table}

\begin{table}[tbh]
\caption{Comparison of the moments  between the simulations and theory  for fBm in $Q=6$}
\begin{center}
\begin{tabular}{|l|c|c|c|c|c|c|c|c|c|}
\multicolumn{4}{c}{}\\ \hline
$H$& 0.1 &0.2&0.3&0.4&0.5&0.6&0.7&0.8&0.9\\ 
 \hline \hline
$N=512, \mu_2$&
1.2130	&	1.1989	&	1.1846	&	1.1723	&	1.1663	&	1.1787	&	1.2603& 1.7062&3.9633\\
\hline 
$N=64, \mu_2$&
1.2105	&	1.1964	&	1.1827	&	1.1705	&	1.1647	&	1.1743	&	1.2274& 1.3901&1.8394\\
\hline
Theory&
1.227	&	1.206	&	1.188	&	1.173	&	1.1667	&	1.180	&	1.290& N.A.& N.A.
\\
\hline
\end{tabular}
\label{FBM3}
\end{center}
\end{table}

\section{Appendix D  Properties of the financial time series data}
Table \ref{table1} shows the properties of the data  in  section I \cite{B}.
We  chose   six characteristic    time series  from \cite{Ka} and 
showed the properties of these time series.
Three of them are FX, two  are commodities, and one is a stock index.
As mentioned in   section I, we chose three FX time series
  that  have   large temporal correlations.
  Compared to stock  and bond prices, FX has a strong temporal  correlation, meaning that they swing in one direction and tend to stay that way.
The temporal correlation is  sometimes the trend   created by the central banks and  governments.

\begin{table}[h]
\centering
  \caption{Data spec of Table \ref{table1} where USD is US Dollar. CAD is Canadian Dollar, EUR is Euro,  GBP is Britain Pound, Soy  is   the commodity  price of beans,  VIX is the volatility index, and NKY 225 is the   index of Japanese stock market. }
  \begin{tabular}{|c|c||c|c|c|c|}  \hline
 no. &   data name & data category & data period & data interval &data length \\ \hline \hline
1&USD/CAD & FX (natural resource) & 2021/5/19 20:22-2021/7/15 19:57 & minutely & 60,000 \\
\hline
2&EUR/CHF & FX (cross pair) & 2021/5/18 13:20-2021/7/15 20:00 & minutely & 60,000 \\
\hline
3&EUR/GBP & FX (cross pair) & 2021/5/19 1:06-2021/7/15 20:00 & minutely & 60,000 \\
\hline
4&SOY  & commodity & 2021/4/19 9:01-2021/7/15 19:57 & minutely & 38,483 \\
\hline
5&VIX & commodity & 2021/4/15 20:11-2021/7/15 19:56 & minutely & 48,106 \\ \hline
6&NKY225 & stock index & 1965/1/5-2021/7/15 & daily & 14,886 \\
\hline
  \end{tabular}
\label{table1}
\end{table}
\section{Appendix E  Perturbation of the Wishart matrices}
In this Appendix, we consider the 
moments  of perturbation of the Wishart matrices.
We consider  the following matrix,  
\begin{equation}
(A_{\mu},A_{\mu+L},\cdots,A_{(N-1)\mu+L}) =\xi^{\mu}\bm 
 {B}/\sqrt{b^2+1}+\bm{\epsilon^{\mu}}/\sqrt{b^2+1},
 \label{pmc}
\end{equation}
where $\mu=1,2, \cdots, L$ and $\bm{B}$, and $\bm{\epsilon}$ are the size $N$ vectors.
$A_i$ is the element of Eq.(\ref{EX}).
The elements of $\bm{\epsilon} $  and
$\xi^{\mu}$  are i.i.d.  All elements of $\bm{B}$  are $b$.
The first term  corresponds to the perturbation from the random matrix.
The second moment of the eigenvalue  is 
\begin{eqnarray}
    \mu_2&=&\frac{1}{L^2N}\sum_{\nu_1=1}^L\sum_{\nu_2 
 =1}^L\sum_{m_1 =1}^N
    \sum_{m_2 =1}^N< A_{m_1 \nu_1}^T A_{\nu_1 m_2}A_{m_2 \nu_2}^T A_{\nu_2 m_1}> \nonumber \\
    &=&
    1+\frac{1}{Q}+(\frac{b^2}{b^2+1})^2\frac{N(N-1)L^2}{N^3L^2} 
    =
    1+\frac{1}{Q},
\end{eqnarray}
in the limit of $N, L \rightarrow \infty$ with $L/N=Q$.
The second moment is the  same as MPD.
In the same way we can calculate the higher moments which are the  same as MPD.
In this model the correlation is smaller than  that in our model.
These conclusions are consistent with 
 the following  transition of the largest eigenvalue in  this model \cite{pca}. 
 The transition point is 
$Q b_c^4=1$.
Here we set the eigenvector of the largest eigenvalue
as  $\bm{e}_1$.
The order parameter  of the phase transition is 
$|\hat{m}|=\bm{e}_1^{T} \cdot \bm{B}/N$ which is
 the direction cosine.
The critical exponent is,
\begin{equation}
|\hat{m}|\propto  t^{\beta}=((b^2-b_c^2)/b_c^2)^{\beta} \hspace{2cm}{\rm at} \hspace{2cm}b^2\sim b_c^2
\end{equation} where  $\beta=1/2$ and  
\begin{equation}
    \xi_{\infty}\propto t^{-\nu}=t^{-3} \hspace{0.8cm}{\rm at} \hspace{2cm}b^2\sim b_c^2,
\end{equation}
where $\xi_{\infty}$ is the correlation length \cite{pca}.

\section{Appendix F  Correlated time series}
\label{AF}
In this Appendix  we review how to create the correlated time series.
We create 
$\bm{A}_\mu$ from the non-correlated normal distribution, $\bm{A}_{0\mu}$ where
\begin{equation}
A_0= (\bm{A}_{01}, \bm{A}_{01+L}, \cdots, \bm{A}_{01+(N-1)L}),  
\label{X0}
\end{equation}
and  $\bm{A}_{0\mu}= (A_{0\mu},\cdots, A_{0\mu+L-1})^T$,
$\mu=1, 1+L, \cdots, 1+(N-1)L$,  the size $L$ vector.

The distribution $\bm{A}_{\mu}$ is the correlated normal distribution,
\begin{equation}
f(\bm{A}_\mu)=\frac{1}{(2\pi)^{\frac{L}{2}}|D_{L-1}|}
\exp[-\frac{1}{2}\bm{A}_\mu^T D_{L-1}^{-1} \bm{A}_\mu].
\end{equation}

Here we  set
\begin{equation}
  \bm{A}_\mu=\sqrt{\Pi} \bm{A}_{0\mu}.
\end{equation}

The distribution of $\bm{A}_{0\mu}$,  $g(\bm{A}_{0\mu})$ is 
\begin{equation}
g(\bm{A}_{0\mu})=f(\bm{A}_\mu)|\frac{\bm{A}_\mu}{\bm{A}_{0\mu}}|
=\frac{|\sqrt{\Pi}|}{{(2\pi)^{\frac{L}{2}}|D_{L-1}|}}
\exp[-\frac{1}{2}\bm{A}_{0\mu}^T\sqrt{\Pi} ^T D_{L-1}^{-1} \sqrt{\Pi}\bm{A}_{0\mu}].
\end{equation}
When  $\sqrt{\Pi}^T D_{L-1}^{-1} \sqrt{\Pi}=\bm{1}$,
the distribution, $g(\bm{A}_{0\mu})$ is 
\begin{equation}
g(\bm{A}_{0\mu})=\frac{1}{(2\pi)^{\frac{L}{2}}|D_{L-1}|}
\exp[-\frac{1}{2}\bm{A}_{0\mu}^T \bm{A}_{0\mu}],
\end{equation}
 and Eq.(\ref{pi}).
 This is the non-correlated normal distribution.

\end{document}